\documentclass[twocolumn]{aastex62}

\usepackage{graphicx}
\usepackage{txfonts}
\usepackage{multirow}
\usepackage{array}

\newcommand{\dn}{DN\,s$^{-1}$\,px$^{-1}$}

\begin{document}

\title{Manifestations of three dimensional magnetic reconnection in an eruption of a quiescent filament: \\ Filament strands turning to flare loops}

\correspondingauthor{J. L\"{o}rin\v{c}\'{i}k}
\email{juraj.lorincik@asu.cas.cz}

\author[0000-0002-9690-8456]{Juraj L\"{o}rin\v{c}\'{i}k}
\affil{Institute of Astronomy, Charles University, V Hole\v{s}ovi\v{c}k\'{a}ch 2, CZ-18000 Prague 8, Czech Republic}
\affil{Astronomical Institute of the Czech Academy of Sciences, Fri\v{c}ova 298, 251 65 Ond\v{r}ejov, Czech Republic}

\author[0000-0003-1308-7427]{Jaroslav Dud\'{i}k}
\affil{Astronomical Institute of the Czech Academy of Sciences, Fri\v{c}ova 298, 251 65 Ond\v{r}ejov, Czech Republic}

\author[0000-0001-5810-1566]{Guillaume Aulanier}
\affil{LESIA, Observatoire de Paris, Universit\'e PSL , CNRS, Sorbonne Universit\'e, Universit\'e Paris-Diderot, 5 place Jules Janssen, 92190 Meudon, France}

\begin{abstract} 
We report on observations of conversion of bright filament strands into flare loops during 2012 August 31 filament eruption. Prior to the eruption, individual bright strands composing one of the legs of the filament were observed in the 171\,\AA{} filter channel data of the Atmospheric Imaging Assembly. After the onset of the eruption, one of the hooked ribbons started to propagate and contract, sweeping footpoints of the bright filament strands as well as coronal loops located close by. Later on, hot flare loops appeared in regions swept by the hook, where the filament strands were rooted. Timing and localization of these phenomena suggest that  {they are caused by reconnection of field lines composing the filament at the hook, which, to our knowledge, has not been observed before.} This process is not included in the standard flare model (CSHKP), as it does not address footpoints of erupting flux ropes and ribbon hooks. It has, however, been predicted using the recent three-dimensional extensions to the standard flare model. There, the erupting flux rope can reconnect with surrounding coronal arcades as the hooked extensions of current ribbons sweep its footpoints. This process results in formation of flare loops rooted in previous footpoints of the flux rope. Our observations of sweeping of filament footpoints are well described by this scenario. In all observed cases, all of the footpoints of the erupting filament became footpoints of flare loops. This process was observed to last for about 150 minutes, throughout the whole eruption. 
\end{abstract}


\section{Introduction} \label{sec_intro}

Solar flares are characterized by sudden increase of radiation throughout entire electromagnetic spectrum \citep[see e.g.][]{fletcher11}. Many of them are accompanied by eruptions of material into interplanetary space called coronal mass ejections \citep[CMEs, see e.g.][]{schmieder15}. First attempts to simulate CMEs were included in the standard CSHKP model of solar flares \citep{car64,stu66,hir74,kopp76}. In this two-dimensional model, field lines dragged into the magnetic null-point reconnect into flare loops anchored in the chromosphere and field lines escaping the Sun in CME \citep[see e.g.][]{shibata11}. 

CMEs are often observed in the form of eruptions of filaments, or prominences, when observed outside the solar disk \citep[see e.g.][]{schmieder14}. Filaments can be optically thick at EUV wavelengths \citep{anzer05} and therefore are typically observed in the chromospheric lines \citep[see e.g.][]{schmieder17, zapior19}. Numerical models revealed that filaments are located in magnetic dips within large flux ropes composed of twisted field lines \citep[see e.g.][]{aulanier2002,dudik2008,gunar16}.

Eruptions are now simulated in three dimensions. In the standard flare model in three dimensions \citep{aulanier12,aulanier13,janvier13}, the modelled flux rope evolves due to photospheric motions of a bipolar region in which the flux rope is rooted \citep{zucc15}. Upon reaching the threshold of the torus instability \citep{kliemtorok06}, the flux rope undergoes a full eruption. 

Magnetic reconnection accompanying the eruption occurs in quasi-separatrix layers \citep[QSLs,][]{priestdemo95,demopriest96}, in which magnetic connectivity is continuous, but has strong gradients. In QSLs reconnection, neighboring field lines change connectivities gradually as they pass through current layers, which results in apparent slippage of field lines. This phenomenon is reffered as to slipping magnetic reconnection \citep{aulanier06,janvier13,janvier17} and has been observed in numerous flares \citep{dudik14,lizhang14, lizhang15, dudik16,jing17,li18, lorincik19}. 

\begin{figure*}[!t]
\centering
    \includegraphics[width=18cm, clip,  bb= 50 120 885 963]{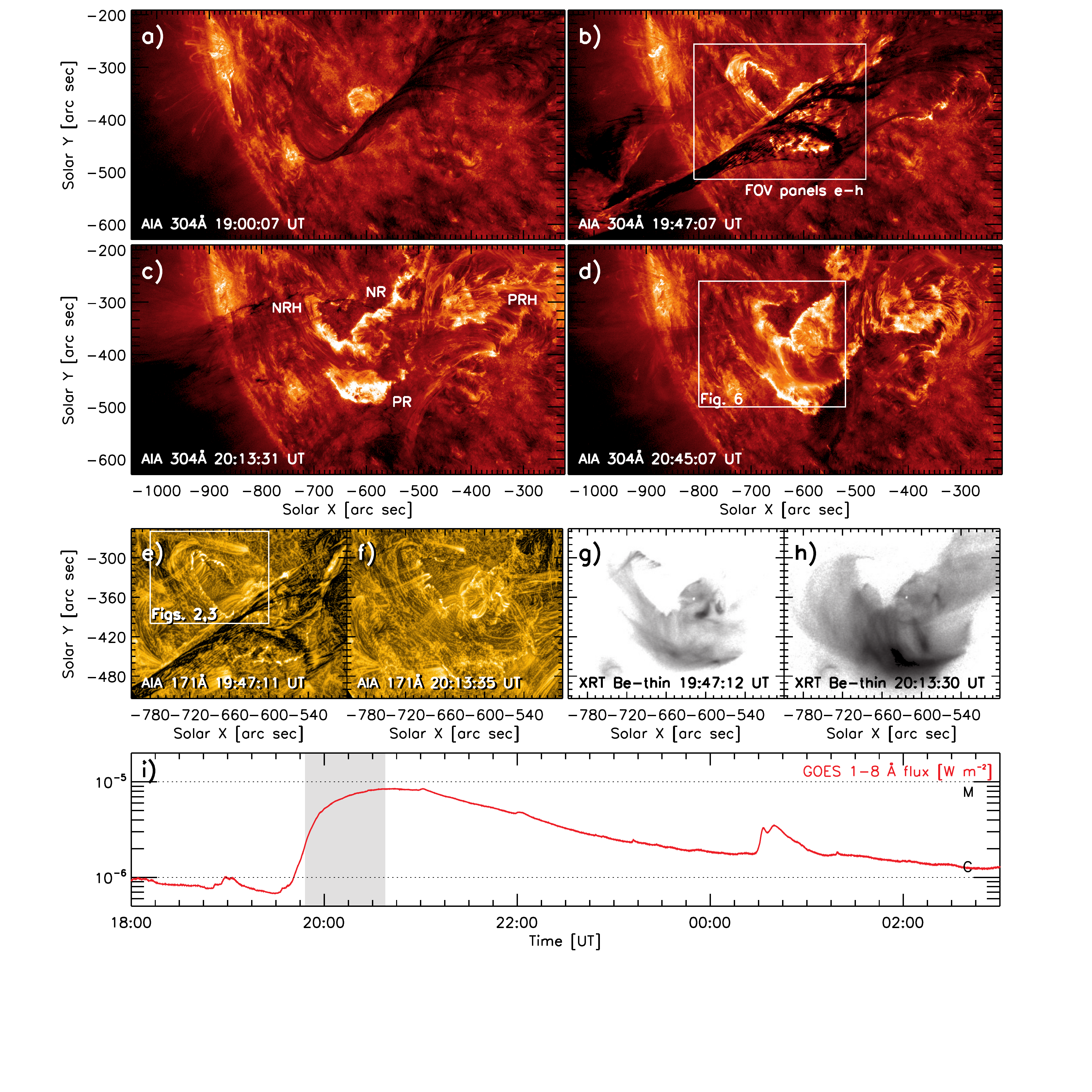}
	\caption{Overview of the filament eruption in the 304\,\AA{} (panels a--d) and 171\,\AA{} (panels e--f) filter channels of \textit{SDO}/AIA. Acronyms NR, NRH, PR, and PRH stand for the negative- and positive-polarity ribbons and ribbon hooks. Panels g--h show \textit{Hinode}/XRT observations carried out using the filter Be-thin. Panel i) shows flux in the 1--8 \,\AA{} channel of \textit{GOES}. Grey shaded area marks period of reconnection of filament strands, as defined in Section \ref{sec_drifting}. Panel a) shows the filament prior to its eruption, panels b), e), and g) show the eruption during the impulsive phase of the flare, and panels c), f), and h) show the peak phase of the flare.
\\ (Animated version of the 171\,\AA{} observations of the eruption is available online.) \label{fig_overview}}
\end{figure*}

Models show that footpoints of flux ropes are encircled by $J$-shaped (hooked) current ribbons \citep{demopriest96, pariat12}, which are intersections of QSLs with the photospherical boundary \citep{aulanier12, janvier13, janvier14b}. These structures correspond to flare ribbons typically observed in the UV \citep{dudik14, janvier16, zhao16, lorincik19}. Observations show that the footpoints of erupting flux ropes are not typically visible, but correspond to coronal dimmings \citep[e.g.][]{dissauer18}.

\citet{aulanier19} used 3D MHD model of erupting flux rope of \citet{zucc15} to investigate different reconnection geometries during eruption of the modelled flux rope with respect to actual positions of the propagating ribbons (reconnection sites). Three types of field lines have been recognized to partake in different reconnection geometries, being 'r' -- flux rope field lines, 'a' -- coronal arcades, and finally 'f' -- flare loops. Based on pre- and post- reconnective field lines partaking in reconnections, three types of magnetic reconnection geometries and sequences accompanying the eruption of flux rope have been recognized:

\begin{itemize}
\item{\textit{aa--rf:}} Two lines from arcade turn into new line composing the flux rope and flare roop. This is a standard reconnection geometry, which is in a simplified 2D form included in the CSHKP model of solar flares.
\item{\textit{rr--rf:}} Two field lines composing the flux rope turn into new flux rope line and flare loop. The flare loop is rooted at straight parts of the QSL footprints along the PIL, while the poloidal flux of the flux rope increases.
\item{\textit{ar--rf:}} Field lines composing the flux rope and inclined arcade turn into new flux rope field line and flare loop. This sequence might be direct via inclined arcade loops, or through an additional reconnection of a low-lying arcade to flux rope and then with another line forming an overlying arcade to a long flare loop (sequential ar--rf reconnection). 
\end{itemize}

The aa--rf and rr--rf reconnections occur below the erupting flux rope, while the ar--rf reconnection occurs at ribbon hooks surrounding flux rope footpoints. It manifests itself in the form of apparent drifting of flux rope footpoints along hooks.

New reconnection geometries with respect to the position of observed ribbons were first investigated by \citet{aulanier19}, who analyzed ribbon hooks in two X-class solar flares observed using the 304\,\AA{} filter channel of AIA or slit-jaw images produced by IRIS. Evolution of the hooks observed during the impulsive phases of both events was consistent with modelled predictions for drifting of footpoints of the erupting flux rope. An observational candidate for the ar--rf reconnection was found in an analysis of flux rope eruption of \citet{zemanova19}. There, one of the ribbon hooks first drifted for many tens of arc seconds, then expanded, and finally shrinked. During the latter two, coronal loops located in a vicinity to the hook became a part of the erupting flux rope and then a flare arcade, which is in agreement with the sequential ar--rf reconnection. However, the erupting flux rope studied in this event is hot (seen in the 131\,\AA{} filter channel of AIA) and its footpoints are only visible for a brief period of time. Therefore, it was difficult to study individual constituents partaking in the ar--rf reconnection.

Here we report on observations of sweeping of filament footpoints by propagating ribbon hook. This was accompanied by fading of a coronal arcade located at the hook and followed by appearance of hot flare loops in swept regions. This paper is structured as follows. In Section \ref{sec_data} we briefly introduce the studied event. Section \ref{sec_observations} details observations of sweeping of filament footpoints and coronal loops. {Formation of arcade of flare loops is described in Section \ref{sec_flareloops}}. We present conclusions in Section \ref{sec_conclusions} and discuss differences between our observations and numerical predictions for this type of reconnection.   

\section{Data} \label{sec_data}

Here we revisit a well-known eruption of a quiescent filament from 2012 August 31, accompanied by a C8.4-class flare. The event was already studied by several authors. For example, \citet{srivastava14} studied partial eruptions of the filament, \citet{williams14} analyzed spectroscopic observations of the eruption, and \citet{wood16} performed a 3D reconstruction of the ejection in the interplanetary space. \citet{lorincik19} investigated motion of flare kernels and apparently slipping flare loops associated with the flare.

To study the eruption we analyze observations from Atmospheric Imaging Assembly \citep[AIA,][]{lemen12} onboard \textit{Solar Dynamics Observatory}. AIA observes the solar corona and flares in six filter channels: 94\,\AA{}, 131\,\AA{} , 171\,\AA{}, 193\,\AA{}, 211\,\AA{}, 335\,\AA{} and solar chromosphere in the 304\,\AA{} channel. Filtergrams in these channels are produced at a cadence of 12 s. AIA also observes the photospheric continumm and the transition region in the 1600\,\AA{} and 1700\,\AA{} filter channels at cadence of 24 s. The spatial resolution of the instrument is $\approx$1.5$\arcsec$, while the pixel size is 0.6$\arcsec$. 

AIA data were processed via the standard \texttt{aia\_prep.pro} routine and corrected for differential rotation. Stray light was deconvolved using the method of \citet{poduval13}. Datasets of the 211\,\AA{}, 335\,\AA{}, 1600\,\AA{}, and 1700\,\AA{} filter channels were then {manually} corrected for {mutual shifts between their coordinate systems. This alignment was performed by matching positions of bright coronal moss and underlying plages.} Compared to the other filter channels of AIA, these shifts were found to equal the size of 1 AIA pixel, i.e. 0.6$\arcsec$ both in X and Y. Finally, The 171\,\AA{} and 211\,\AA{} filter channel data have been processed using the Multi-scale Gaussian normalization (MGN) of \citet{morgan14}. 

In this paper, we also use the measurements of the photospheric magnetic field carried out using the Helioseismic and Magnetic Imager \citep[HMI,][]{scherrer12}. Hot flare emission was reviewed in observations of The X-ray Telescope \citep[XRT,][]{golub07} onboard \textit{Hinode}, carried out using the Be-thin filter. {The XRT data were processed using the standard \texttt{xrt\_prep.pro} and \texttt{xrt\_jitter.pro} routines and then manually co-aligned with EUV filter channels of AIA. To do so we used 131\,\AA{} filter channel observations of $\simeq11$ MK plasma, in which the structure of the hook qualitatively agrees to that observed in the XRT (see, e.g., Figure \ref{fig_overview}g and Figure \ref{fig_nrh_overview}o}).

\begin{figure*}[t]
	\centering
    \includegraphics[width=5.40cm, clip,  viewport= 02 45 283 208]{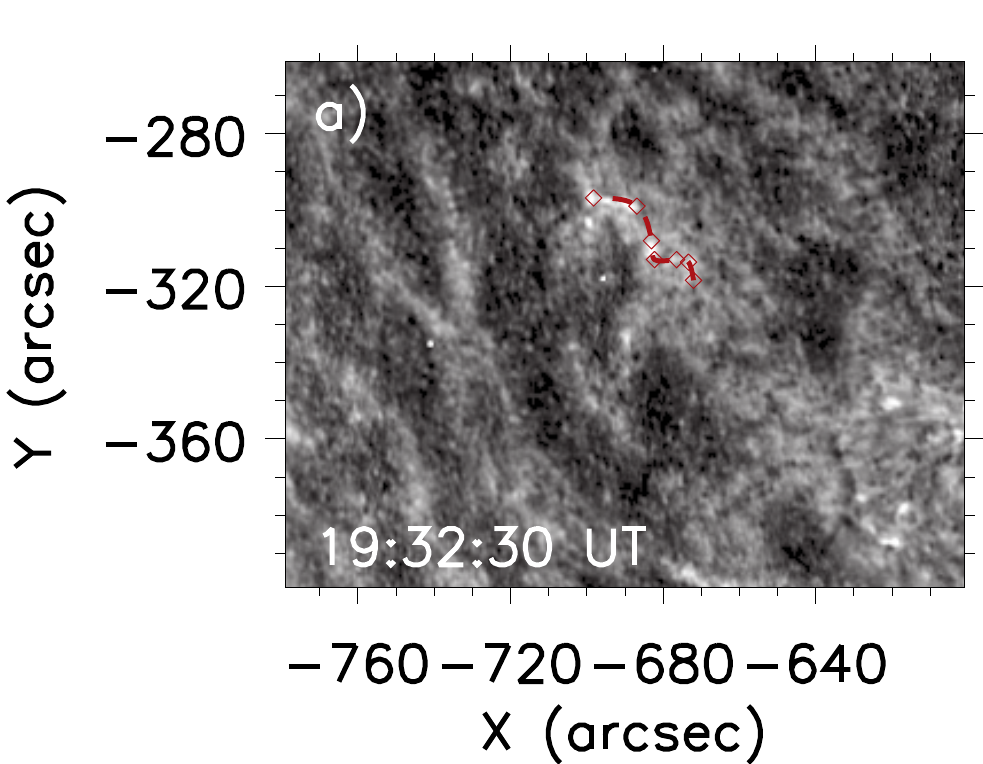}
    \includegraphics[width=3.978cm, clip,  viewport= 76 45 283 208]{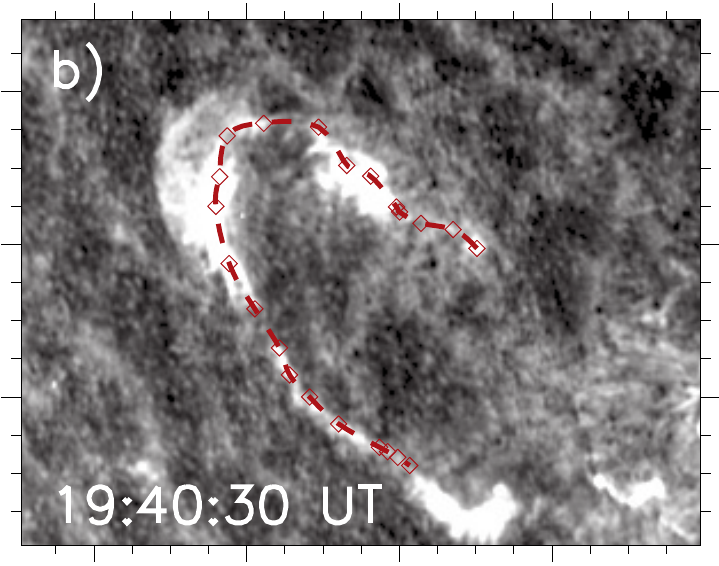}
    \includegraphics[width=3.978cm, clip,  viewport= 76 45 283 208]{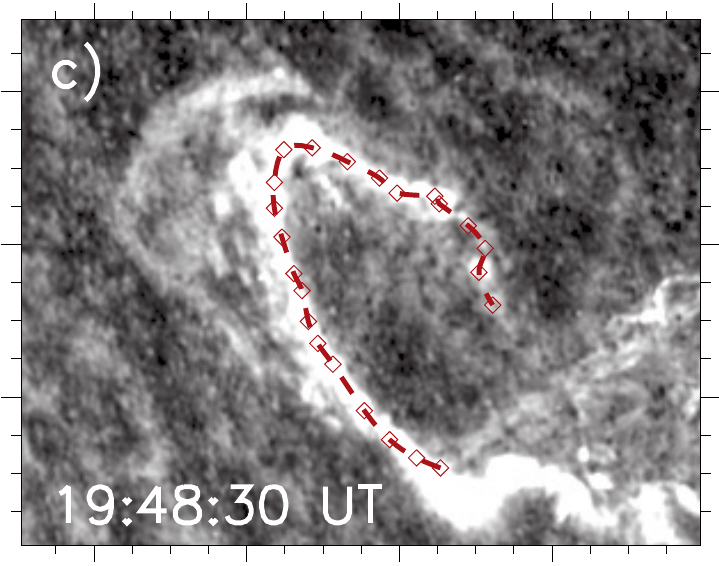}
    \includegraphics[width=3.978cm, clip,  viewport= 76 45 283 208]{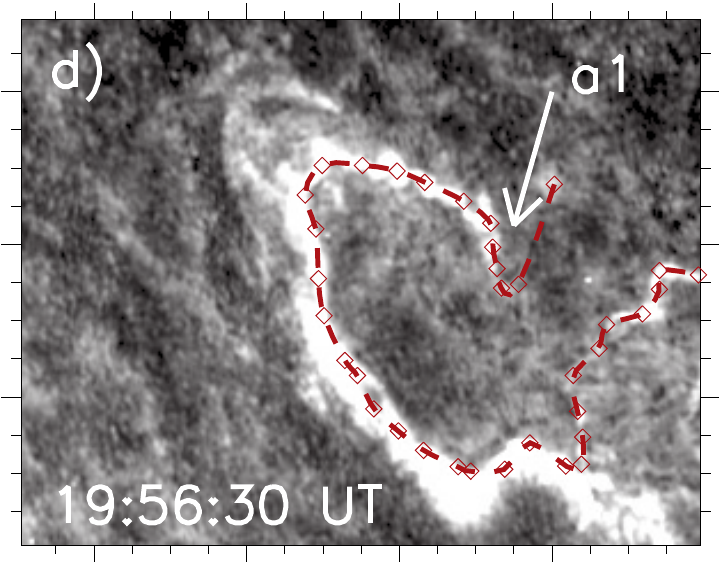}
    	\\
  	\includegraphics[width=5.40cm, clip,  viewport= 02 45 283 208]{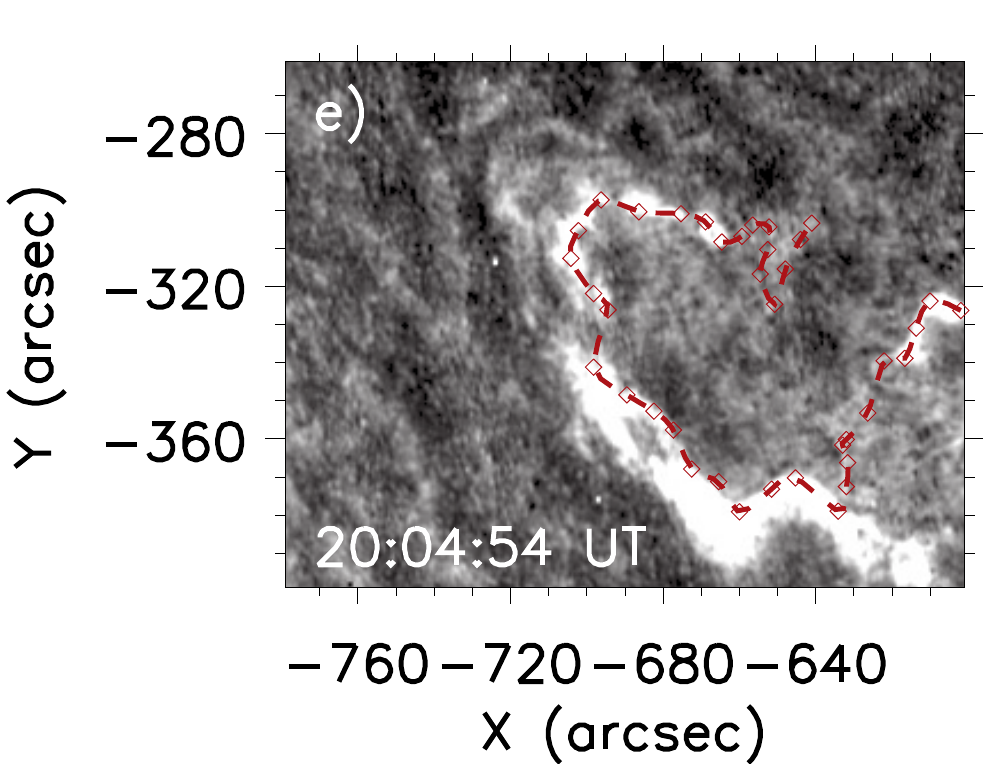}
    \includegraphics[width=3.978cm, clip,  viewport= 76 45 283 208]{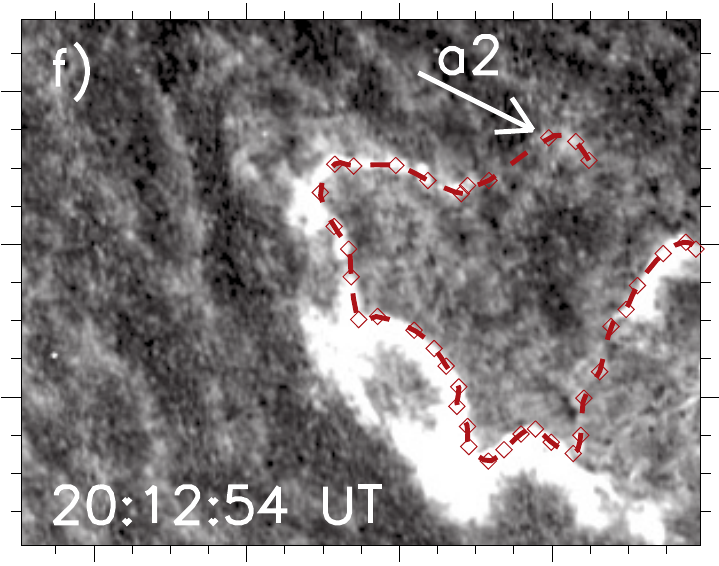}
	\includegraphics[width=3.978cm, clip,  viewport= 76 45 283 208]{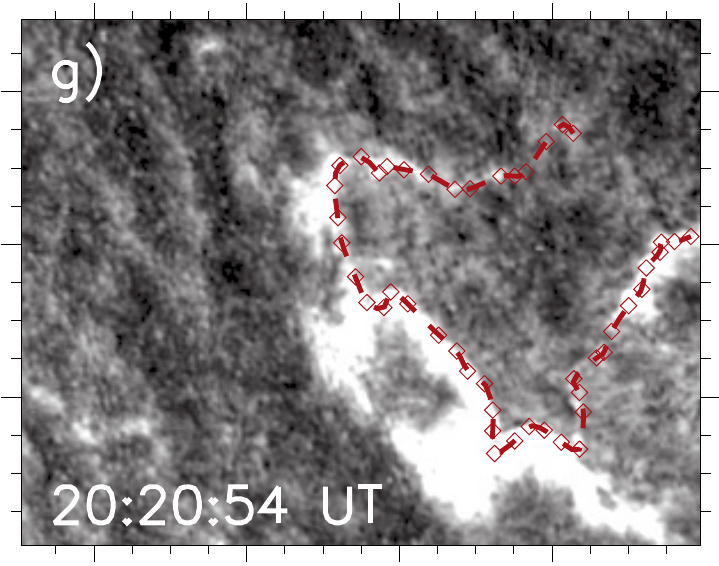}
	\includegraphics[width=3.978cm, clip,  viewport= 76 45 283 208]{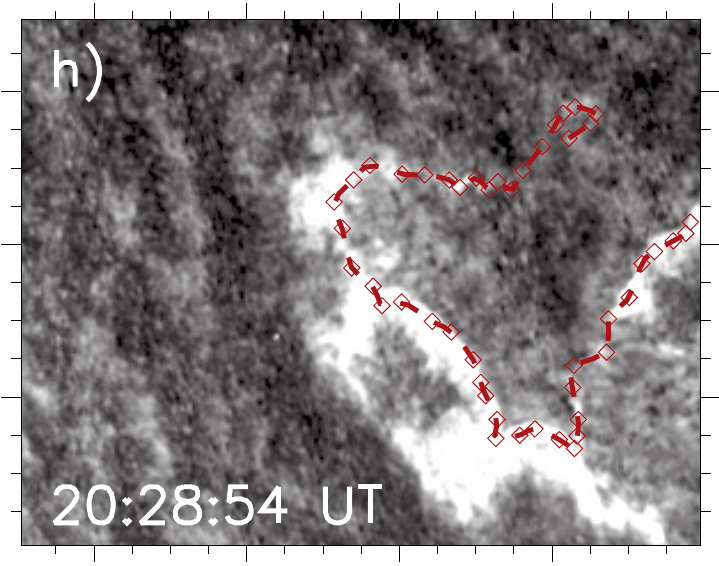}
	\\	
	\includegraphics[width=5.40cm, clip,  viewport= 02 00 283 208]{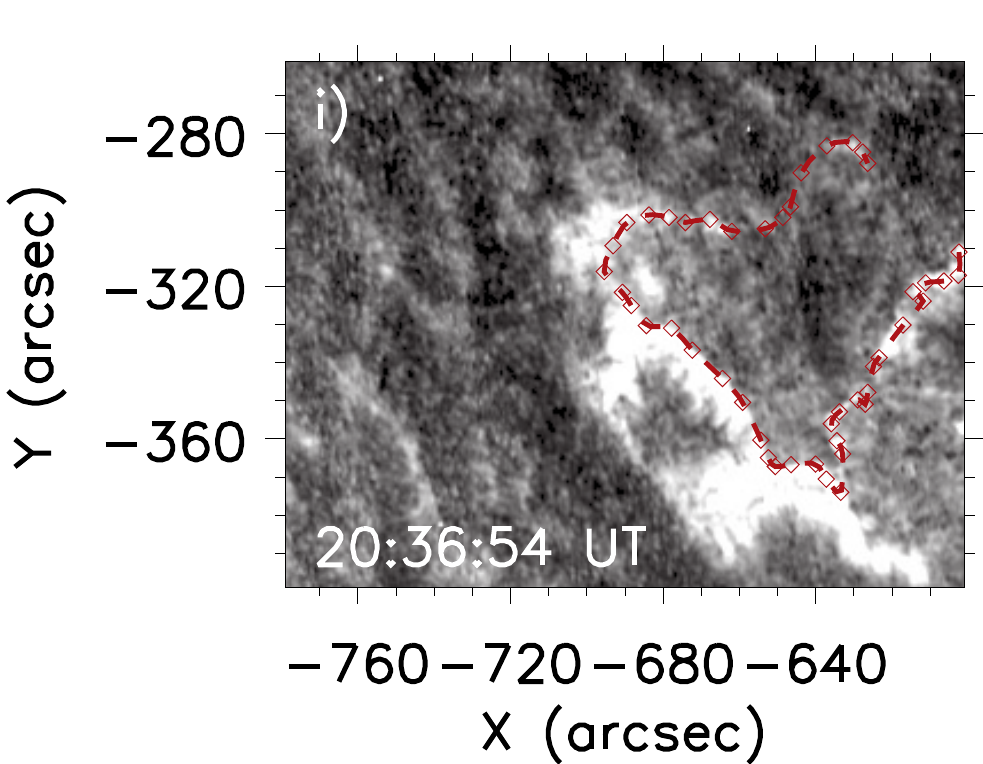}	
	\includegraphics[width=3.978cm, clip,  viewport= 76 00 283 208]{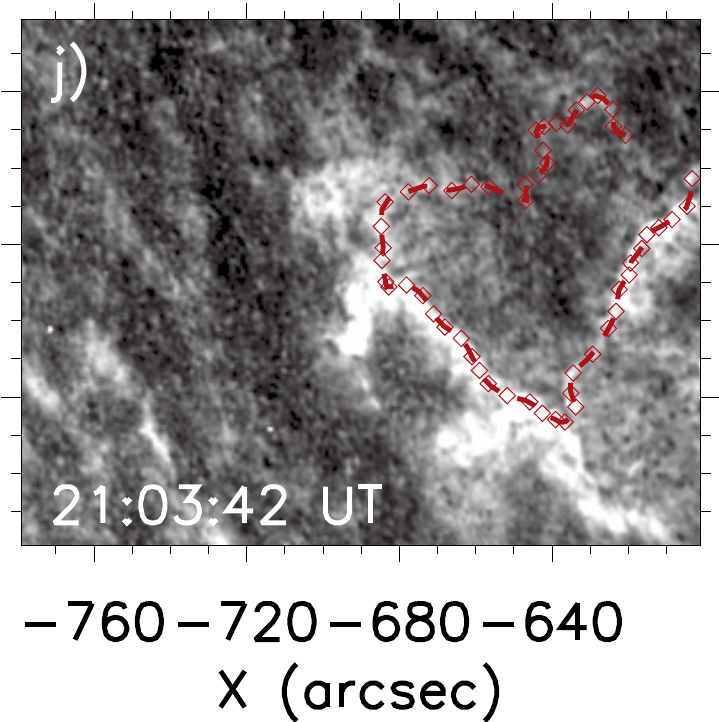}
    \includegraphics[width=3.978cm, clip,  viewport= 76 00 283 208]{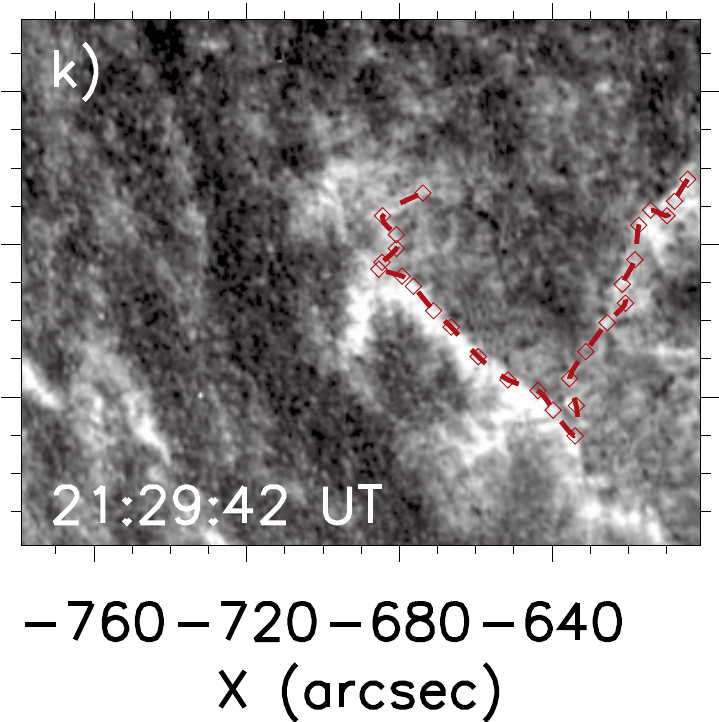}
	\includegraphics[width=3.978cm, clip,  viewport= 76 00 283 208]{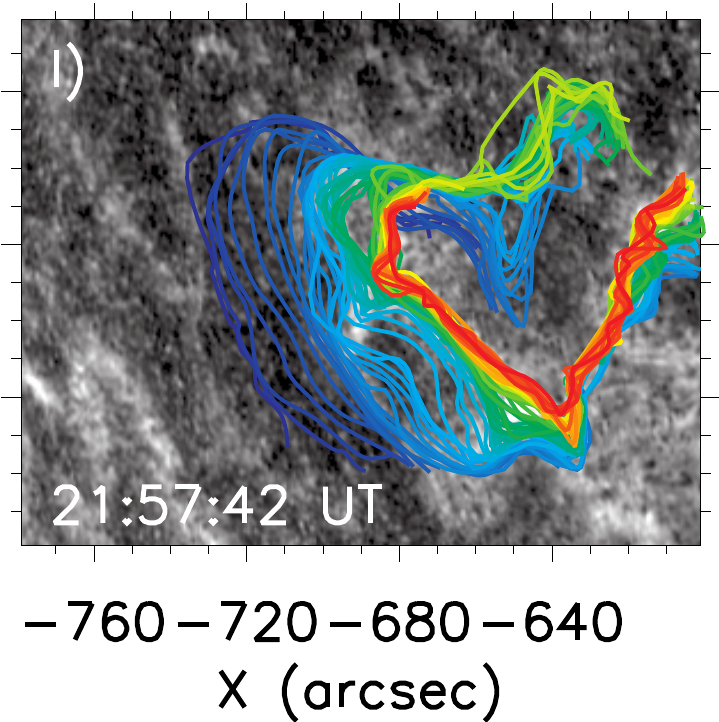}	
	
	\caption{View of NRH in ratios of the 1600\,\AA{} and 1700\,\AA{} filter channels of AIA. Red squares, joined by red curves, mark the adapted positions of the ribbon. Arrows 'a1' and 'a2' in panels d) and f), respectively, point to the tip of NRH further discussed in Section \ref{sec_nrh}. Panel l) shows these positions throughout the eruption. Color-coding was used to distinguish between ribbon positions between the earliest (blue curves, $\approx$19:30 UT) and the latest times (red curves, $\approx$22:00 UT).}
\label{fig_1617_ratio}
\end{figure*}

\section{The filament eruption of 2012 August 31} \label{sec_observations}

\subsection{Overview of the event}

Prior to its eruption on 2012 August 31, the filament was quiescent for more than one Carrington rotation (Figure \ref{fig_overview}a). In Figure \ref{fig_overview}b, two legs of the eruptive filament can be distinguished. The eastern leg was composed of strands anchored in footpoints located in a {close spatial proximity. Some} of the filament strands observed in this leg were bright in all of the EUV filter channels of AIA, and best seen in the 131\,\AA{} and 171\,\AA{} (see Figure \ref{fig_overview}b, e, and the animation accompanying Figure \ref{fig_overview}). The brightness of the filament strands might be due to enhanced TR or coronal emission during the eruption \citep[see e.g.][respectively]{parenti12, williams14}. The western leg was not as spatially concentrated as the eastern leg, as it spread through more than 400$\arcsec$ in the neighborhood of several active regions. Here we also observed several bright strands composing the filament, however, majority of them were obscured by the rest of the erupting filament at some point during the event.


Rising of the filament can first be distinguished in the EUV filter channels data at around 19:30 UT (see the animation accompanying Figure \ref{fig_overview}). At the same time, the \textit{GOES} X-ray emission started to increase (Figure \ref{fig_overview}i). During the eruption of the filament, the \textit{GOES} emission kept rising until it reached its broad maximum at $\approx$20:15--20:30 UT. \textit{Hinode}/XRT data reveal hot flare loops after the onset of the eruption (panel g), with a bulk of emission present underneath the erupting filament. At the maximum of the impulsive phase, the arcade of flare loops was fully formed (panel h). Later, during the gradual phase of the flare (after $\approx$20:30 UT) this arcade progressively cooled, as it was observed in filter channels of AIA with lower typical temperatures, such as 304\,\AA{} (panel d, also see Section \ref{sec_cooling}). Filament apex left the field of field-of-view of the instrument at around 20:00 UT (panels b--c), but its legs remained observable until $\approx$21:30 UT, when they completely disappeared.

A pair of flare ribbons formed at the onset of the eruption. During the course of the eruption, both ribbons underwent complicated elongation and developed hooks. For their description we use acronyms 'N' (resp. 'P') for negative (resp. positive) polarity, 'R' for ribbon, and 'RH' for ribbon hook (Figure \ref{fig_overview}c). 

Panels a--d of Figure \ref{fig_overview} indicate that during the eruption the ribbon hook NRH, encircling the filament's eastern leg, both propagated westward and shrinked. Its morphology and evolution throughout the course of the eruption is detailed in the following section. 

\subsection{Evolution of the negative ribbon hook} \label{sec_nrh}

The evolution of the flare ribbons was studied using data of the 1600\,\AA{} and 1700\,\AA{} filter channels. \citet{dudik16} introduced a method of using ratios of these filter channels in order to enhance the TR emission of \ion{C}{4} in flare kernels with respect to the photospheric contributions. NRH in the observed ratios is shown in Figure \ref{fig_1617_ratio}. In each individual frame, NRH was manually traced by identifying multiple nodes along it (red squares) which were then interpolated using the spline function (red dashed lines). Here we mainly focus on the interaction of the ribbon NR with footpoints of the filament encircled by what later became NRH. Therefore, we were selecting edges of the ribbon NR in the direction of its propagation (ribbon front), while we primarily focused on the morphology of NRH. 

Kernels which composed NR first appeared at $\approx$19:32:20 UT (panel a). The ribbon then elongated to a $J$-shaped structure, creating the hook NRH. Afterwards, NRH started to propagate towards the west and contract. Panels d--e show that the ribbon joined with the kernels located in the internetwork to the west, while the tip of the hook elongated in a $V$-shaped trajectory (arrow 'a1' in panel d). Later, tip of the hook further elongated towards NW in a circular shape (arrow 'a2' in panel f), while the straight part of the ribbon "jumped" over a supergranule located at around \mbox{[--680$\arcsec$,--360$\arcsec$]} (panels f --j). After $\approx$21:00 UT (panels k--l), propagation of the ribbon decelerated and the ribbon started to gradually fade away. 

Panel l) shows the evolution of the ribbon throughout the course of the eruption at a 2-min cadence. Lines indicating the positions of the ribbon were color-coded in orded to distinguish between the times in which the ribbon was observed, $\approx$19:30 UT to $\approx$22:00 UT, coded from blue to red, respectively. Deceleration of the ribbon propagation can be seen by comparing mutual distances between the blue curves at the beginning of the eruption and the red curves at its end. Note the brief acceleration of NRH through a small region at [--690$\arcsec$,--320$\arcsec$] and of the straight portion of NR through the supergranule located at [--680$\arcsec$,--360$\arcsec$].
\begin{figure*}[t]
	\centering
    \includegraphics[width=5.20cm, clip,  viewport= 05 45 281 208]{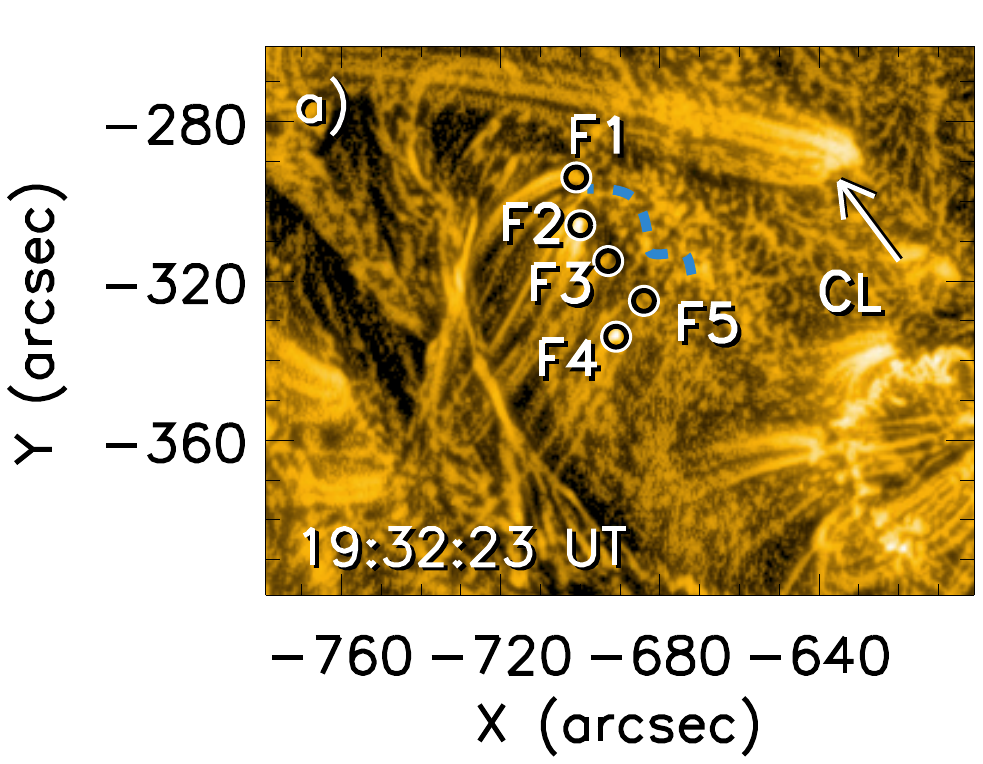}
    \includegraphics[width=3.86cm, clip,  viewport= 76 45 281 208]{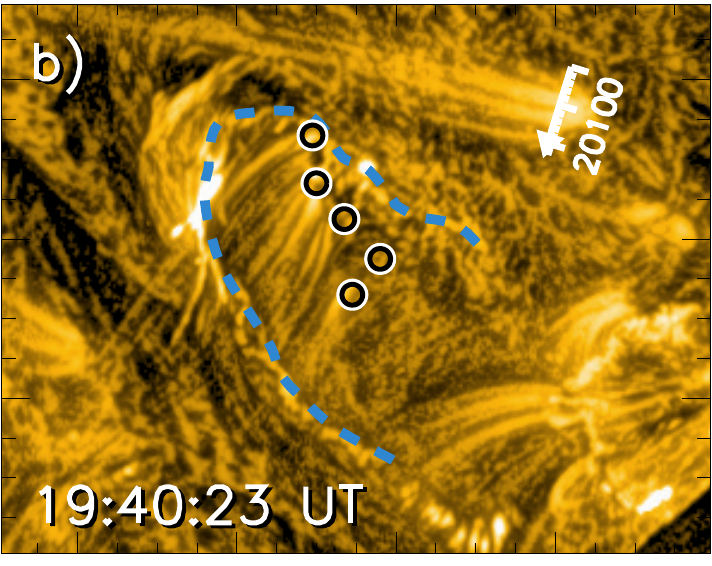}
    \includegraphics[width=3.86cm, clip,  viewport= 76 45 281 208]{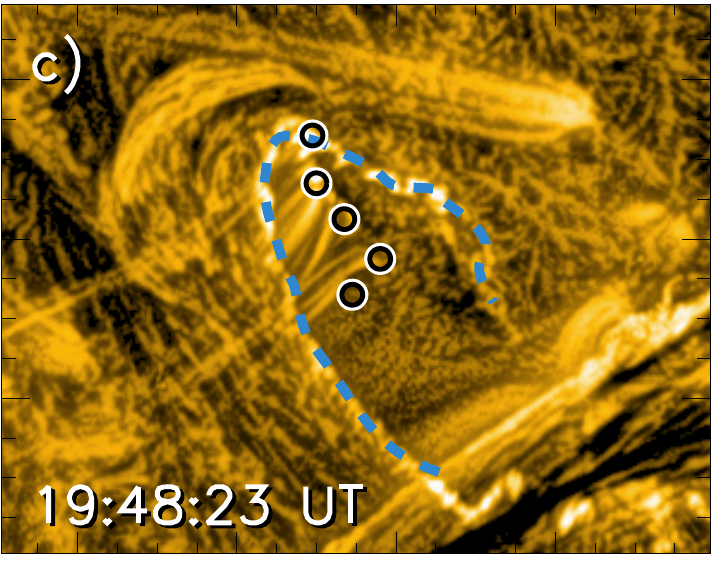}
    \includegraphics[width=3.86cm, clip,  viewport= 76 45 281 208]{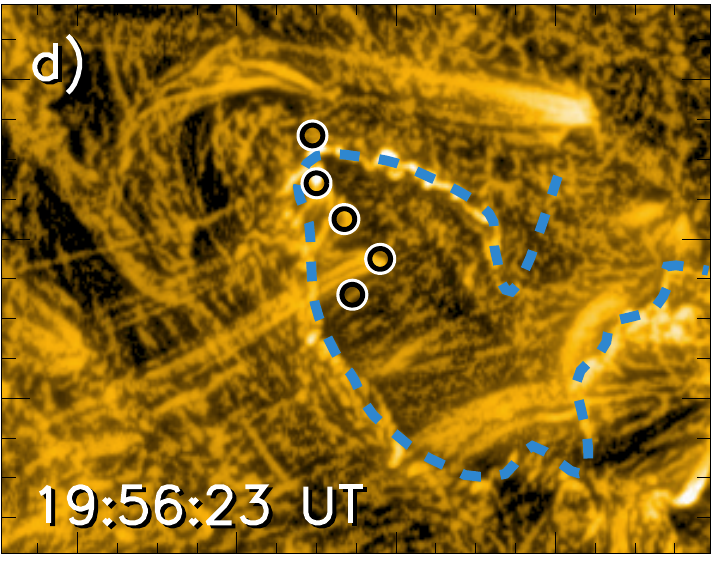}
    	\\
  	\includegraphics[width=5.20cm, clip,  viewport= 05 45 281 208]{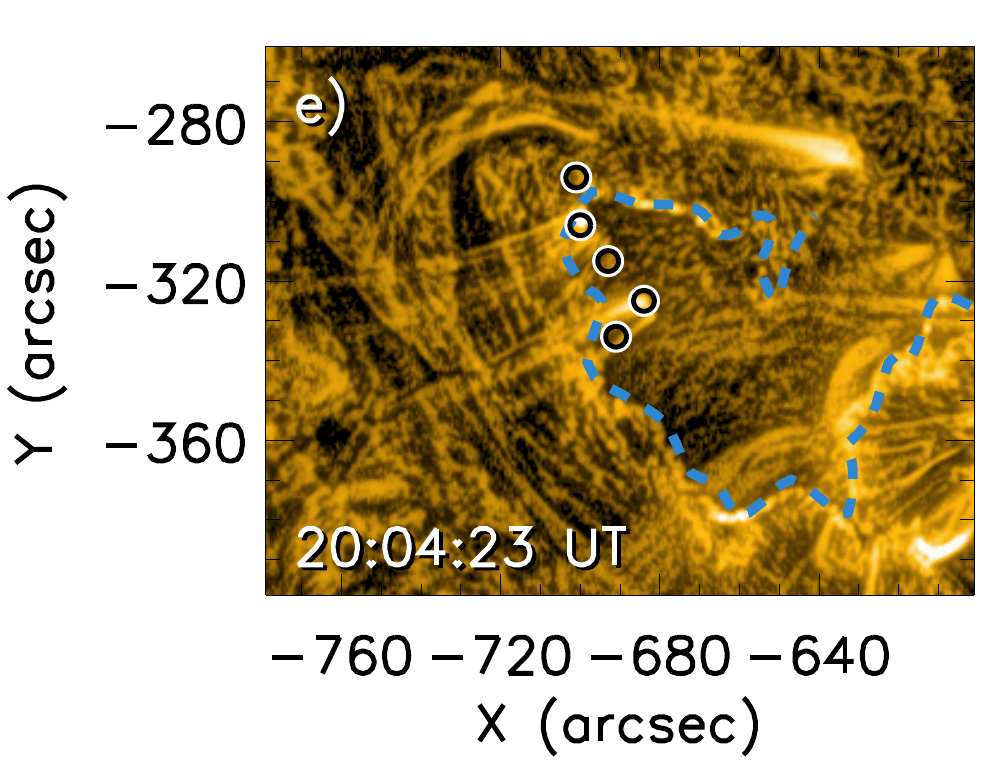}
    \includegraphics[width=3.86cm, clip,  viewport= 76 45 281 208]{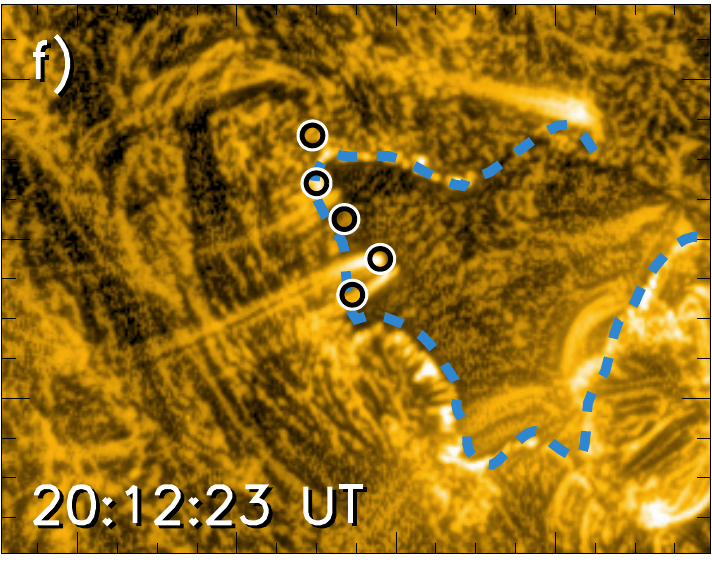}
	\includegraphics[width=3.86cm, clip,  viewport= 76 45 281 208]{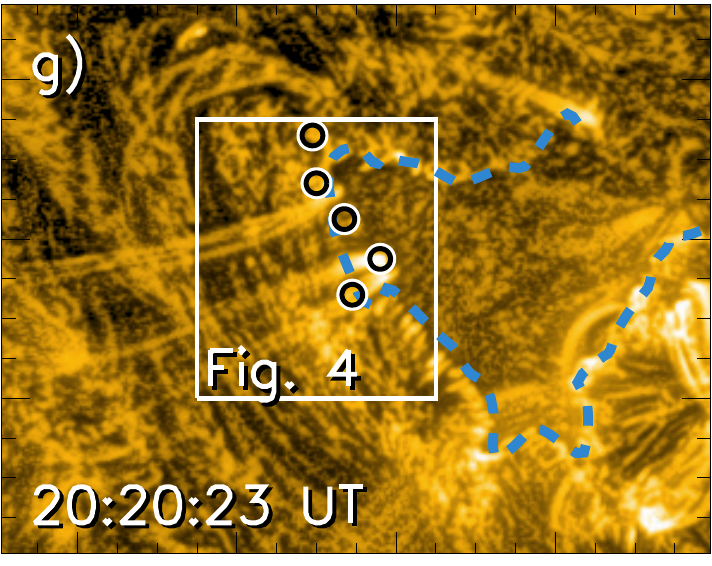}
	\includegraphics[width=3.86cm, clip,  viewport= 76 45 281 208]{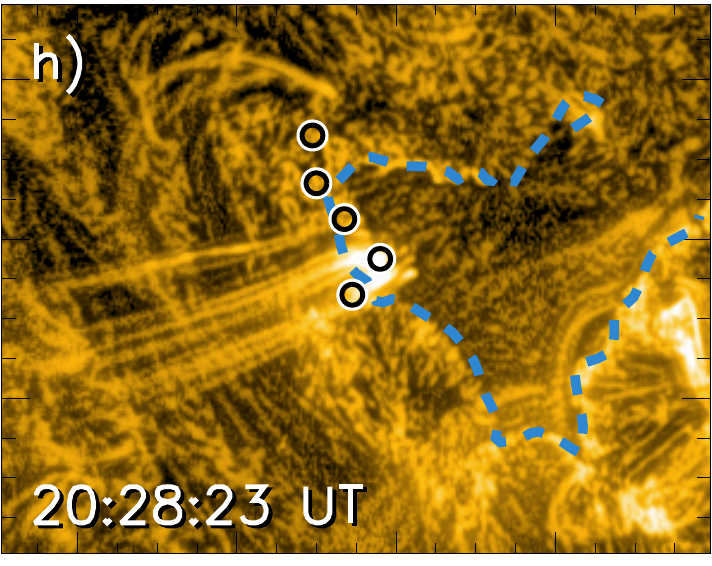}
	\\	
	\includegraphics[width=5.20cm, clip,  viewport= 05 45 281 208]{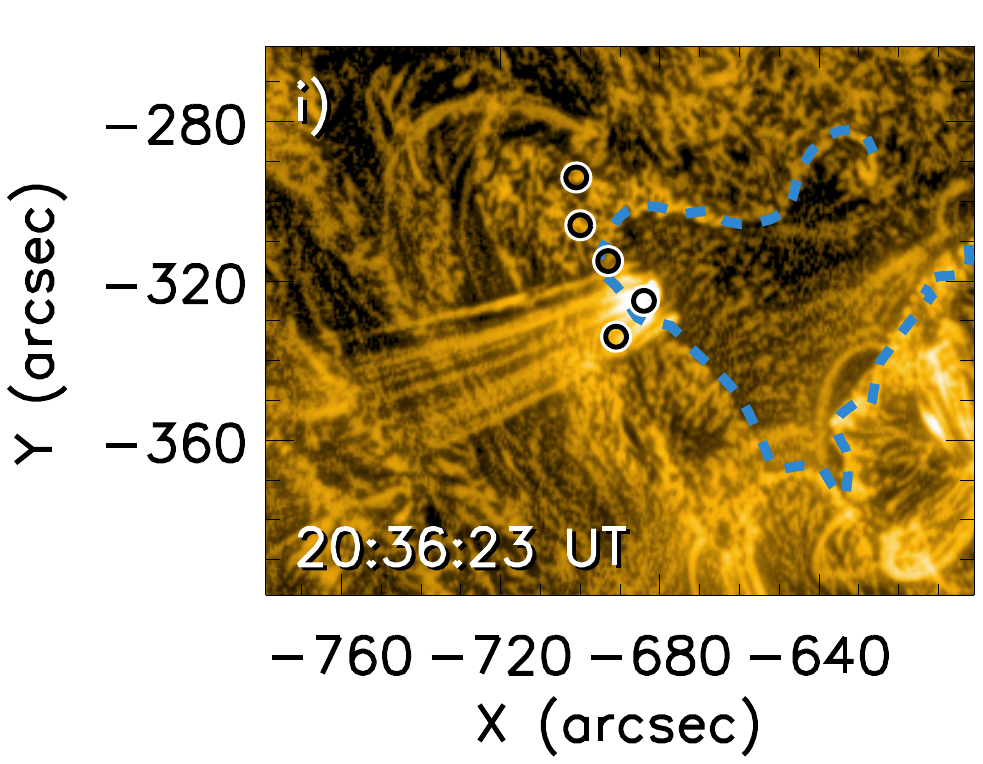}	
	\includegraphics[width=3.86cm, clip,  viewport= 76 45 281 208]{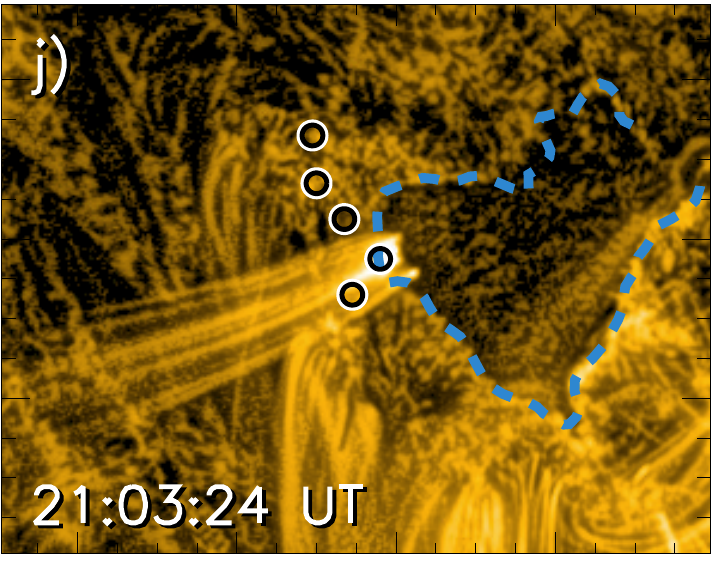}
    \includegraphics[width=3.86cm, clip,  viewport= 76 45 281 208]{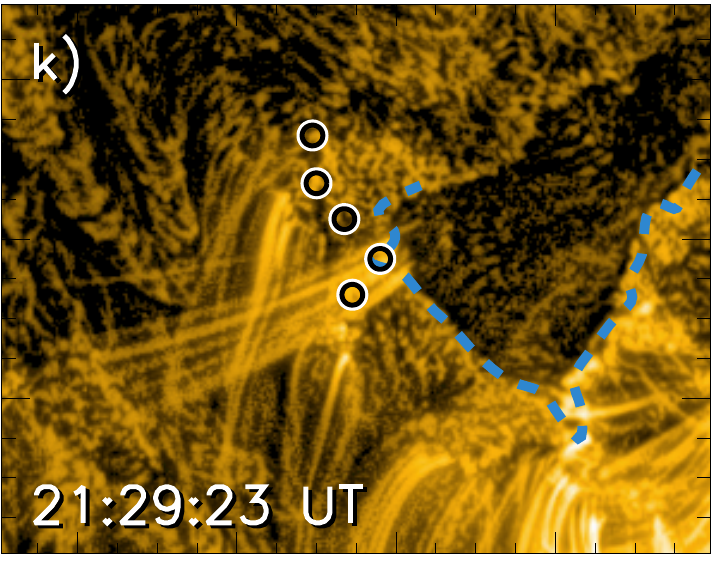}
	\includegraphics[width=3.86cm, clip,  viewport= 76 45 281 208]{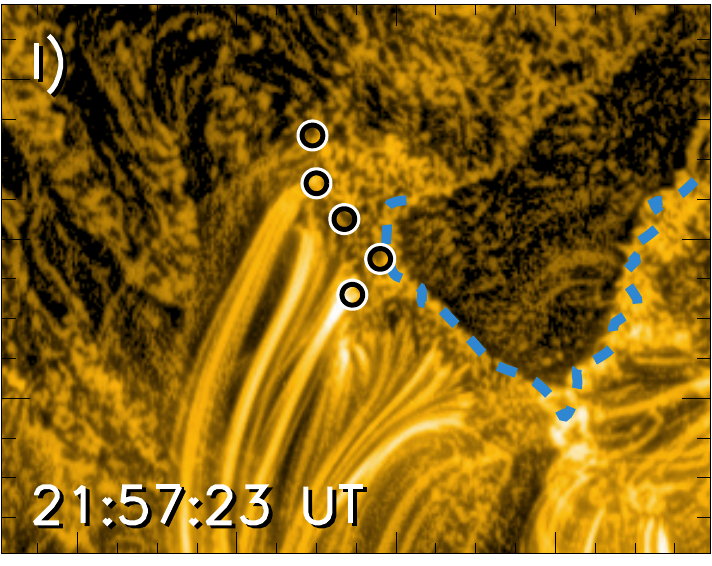}	
	\\
	\includegraphics[width=5.20cm, clip,  viewport= 05 45 281 208]{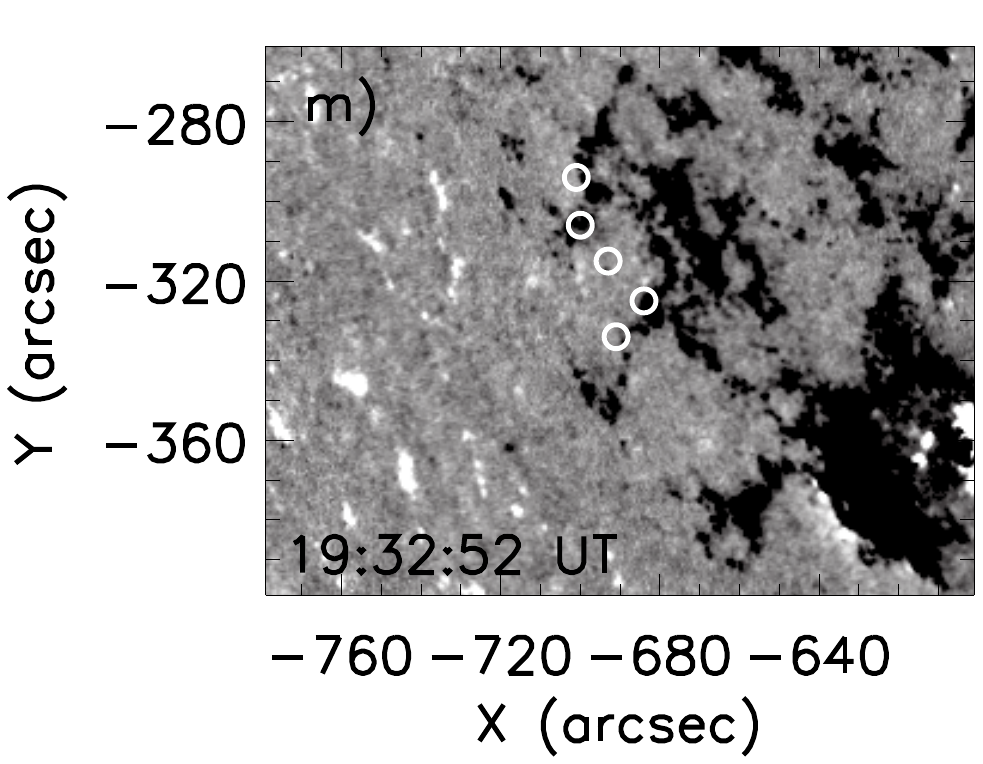}
    \includegraphics[width=3.86cm, clip,  viewport= 76 45 281 208]{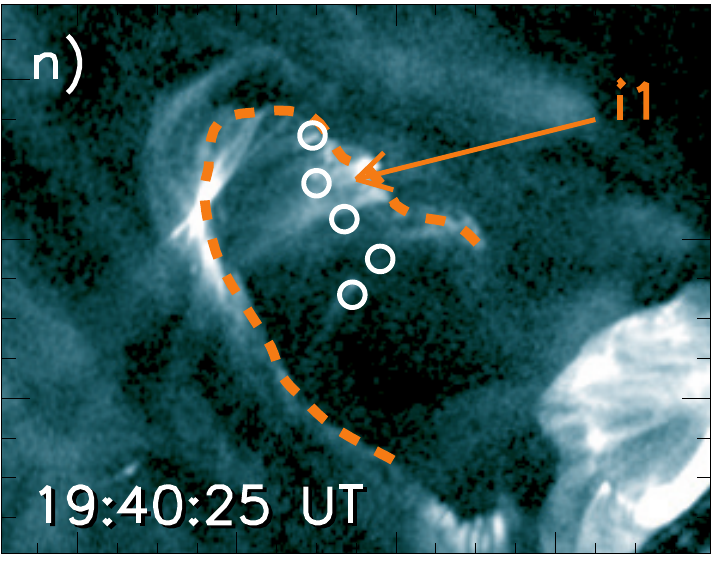}
    \includegraphics[width=3.86cm, clip,  viewport= 76 45 281 208]{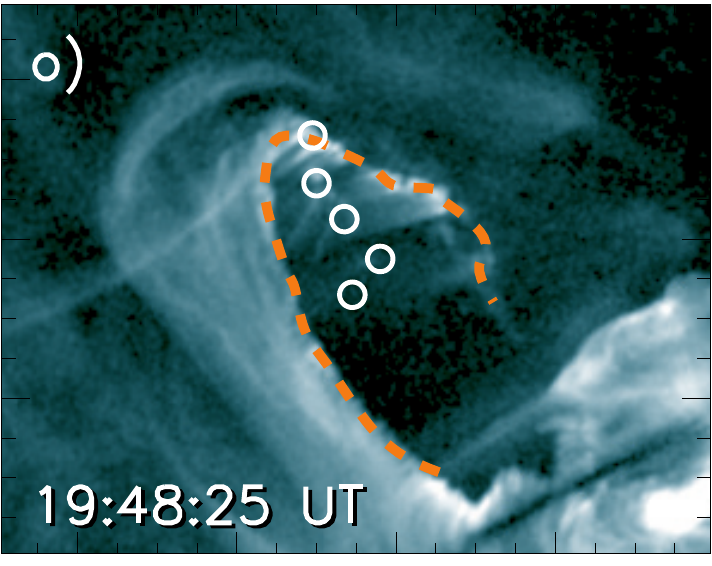}
    \includegraphics[width=3.86cm, clip,  viewport= 76 45 281 208]{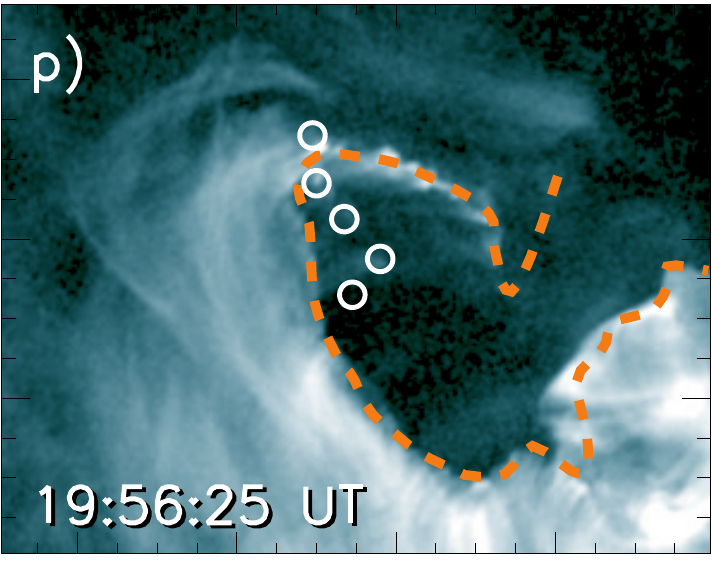}
    \\
	\includegraphics[width=5.20cm, clip,  viewport= 05 45 281 208]{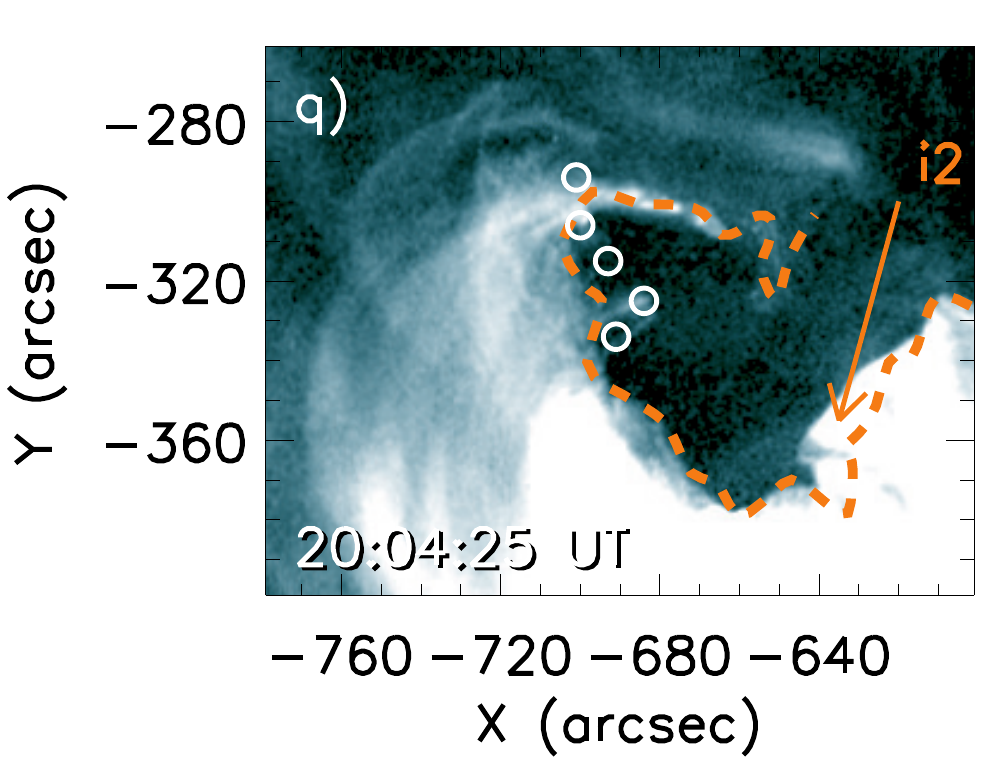}
    \includegraphics[width=3.86cm, clip,  viewport= 76 45 281 208]{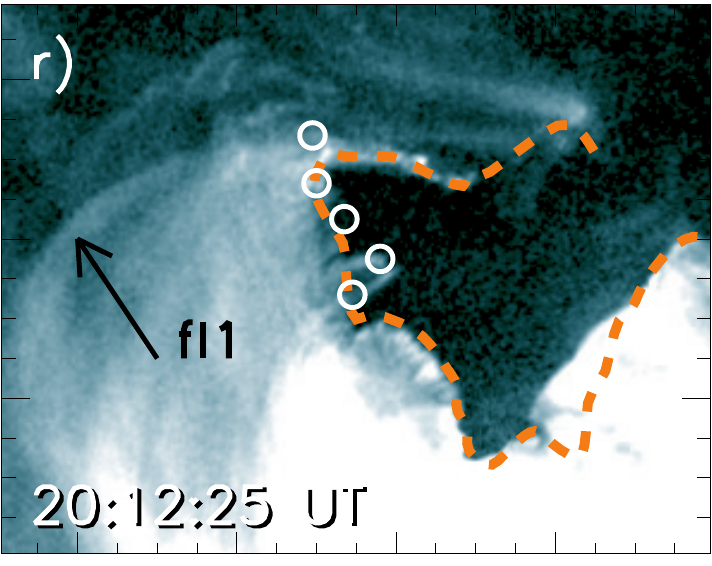}
	\includegraphics[width=3.86cm, clip,  viewport= 76 45 281 208]{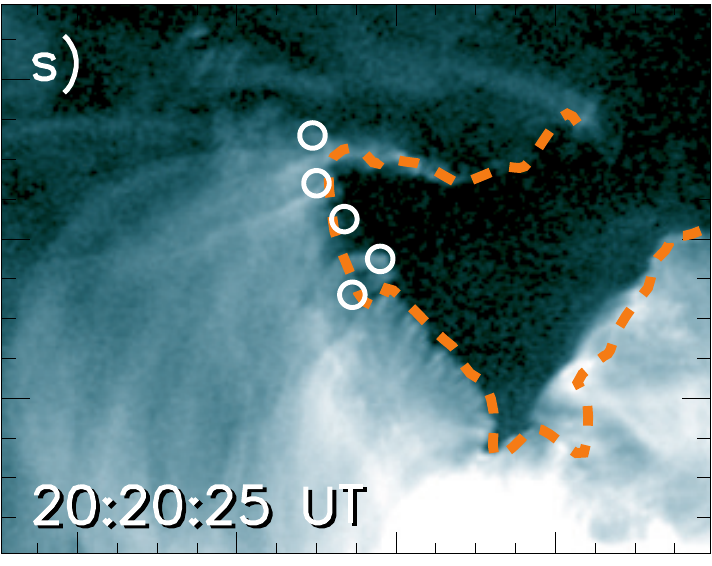}
	\includegraphics[width=3.86cm, clip,  viewport= 76 45 281 208]{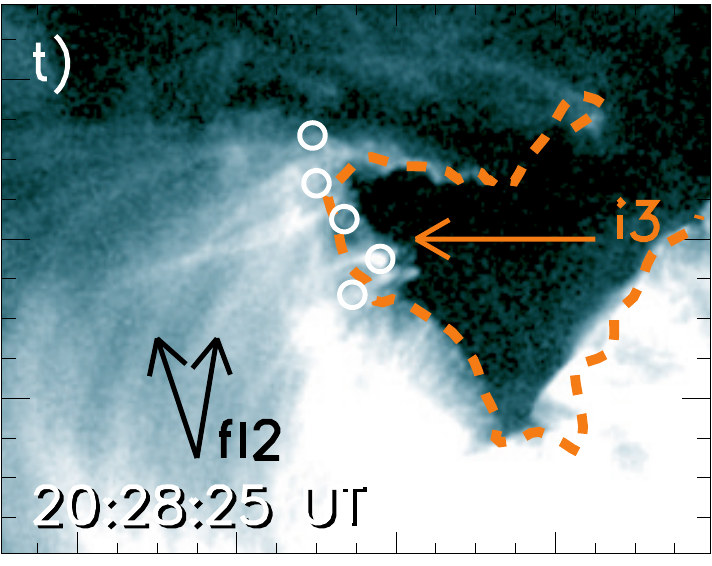}
	\\
	\includegraphics[width=5.20cm, clip,  viewport= 05 05 281 208]{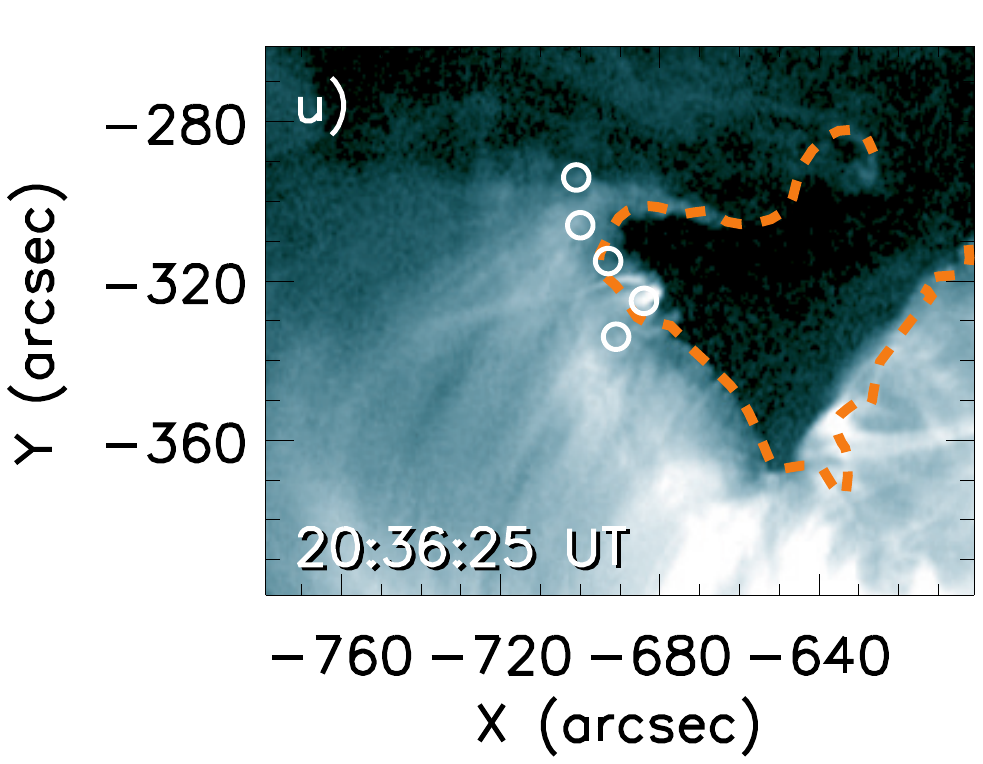}	
	\includegraphics[width=3.86cm, clip,  viewport= 76 05 281 208]{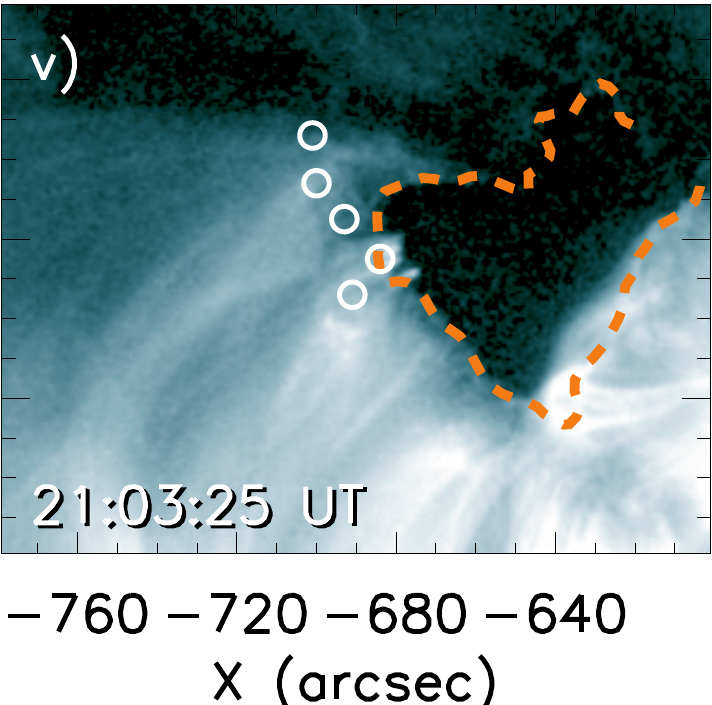}
    \includegraphics[width=3.86cm, clip,  viewport= 76 05 281 208]{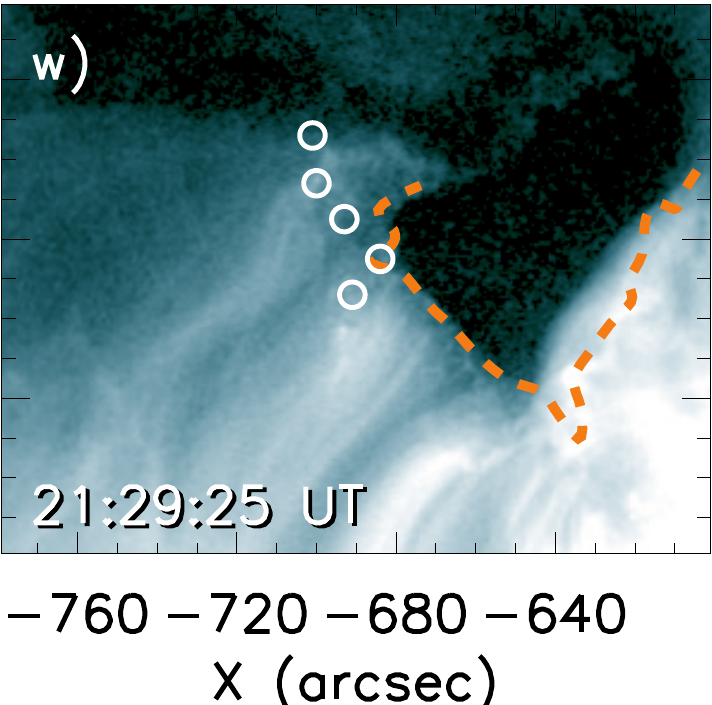}
	\includegraphics[width=3.86cm, clip,  viewport= 76 05 281 208]{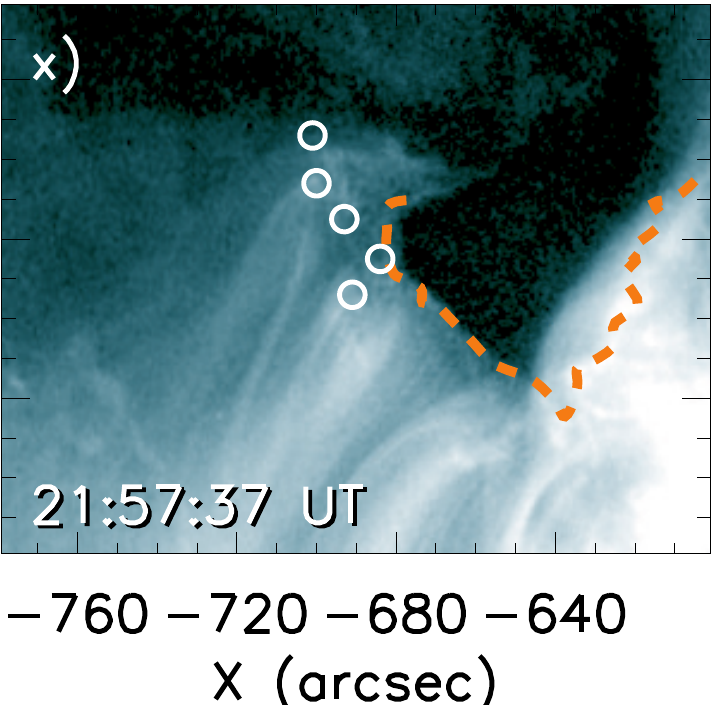}	
	\caption{View of NRH in 171\,\AA{} (panels a -- l) and 94\,\AA{} (panels n -- x) filter channels of AIA. The 171\,\AA{} data shown here are processed using the Multi-Gaussian normalization. Panel m) shows the HMI B$_{\text{LOS}}$ data, saturated to $\pm$ 50 G. White arrow in panel a) marks coronal loops reconnecting with the filament strands. Circles mark footpoints of the filament. Blue and orange dashed lines mark positions of the ribbon found using ratios of the 1600\,\AA{} and 1700\,\AA{} filtergrams (Figure \ref{fig_1617_ratio}). White arrow in panel b) denotes the cut, along which the time-distance diagram shown in Figure \ref{fig_xt_cl} was constructed. Orange arrows i1--i3 highlight emission observed inside NRH. Black arrows fl1 and fl2 indicate individual flare loops within the arcade. \\ (Animated versions of the 94\,\AA{} and 171\,\AA{} observations are available online.) }
\label{fig_nrh_overview}
\end{figure*}

\begin{figure*}[t]
	\centering
    \includegraphics[width=4.9cm, clip,  viewport=  00 90 278 300]{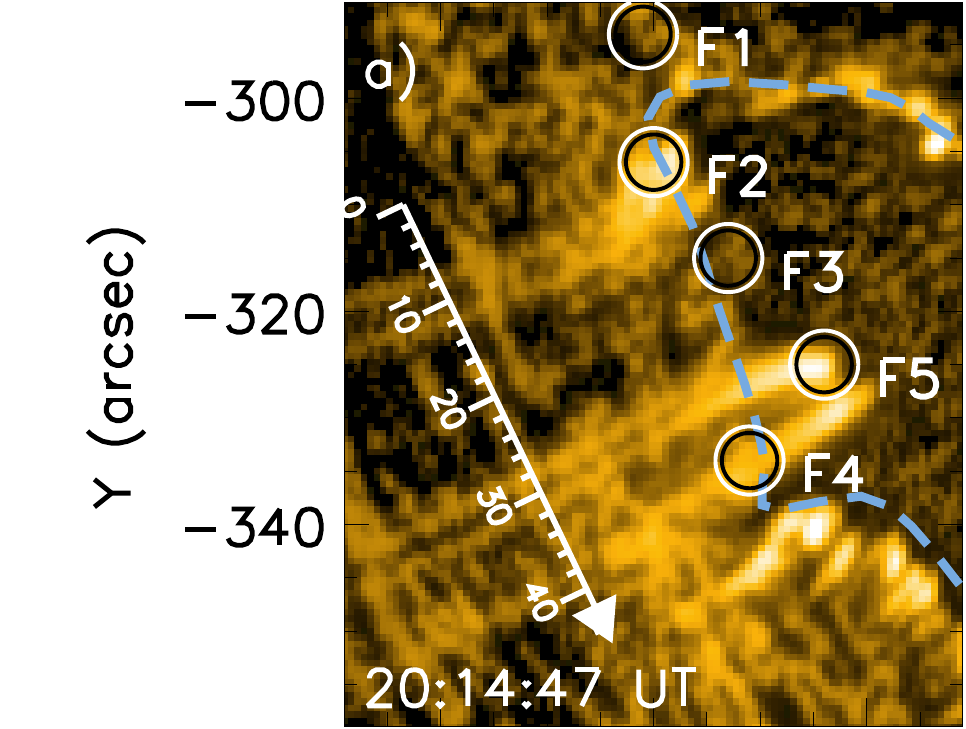}
    \includegraphics[width=3.173cm, clip,  viewport=  98 90 278 300]{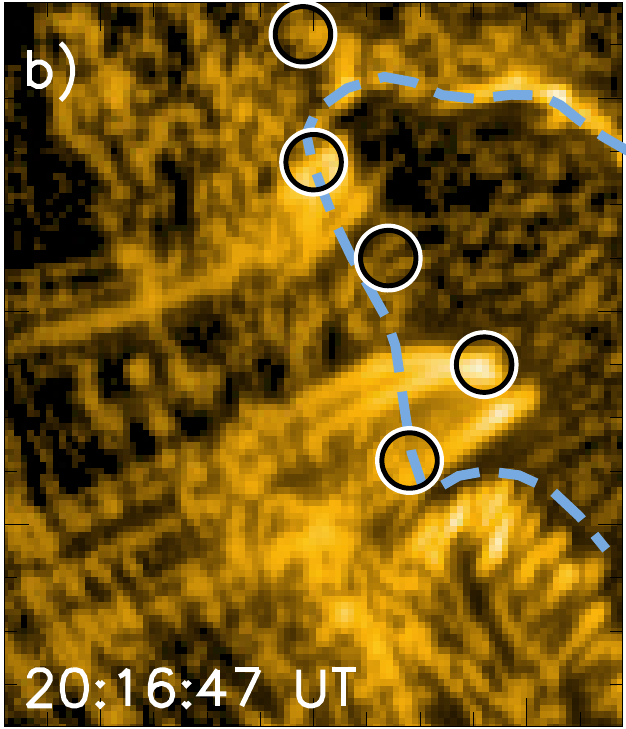}
    \includegraphics[width=3.173cm, clip,  viewport=  98 90 278 300]{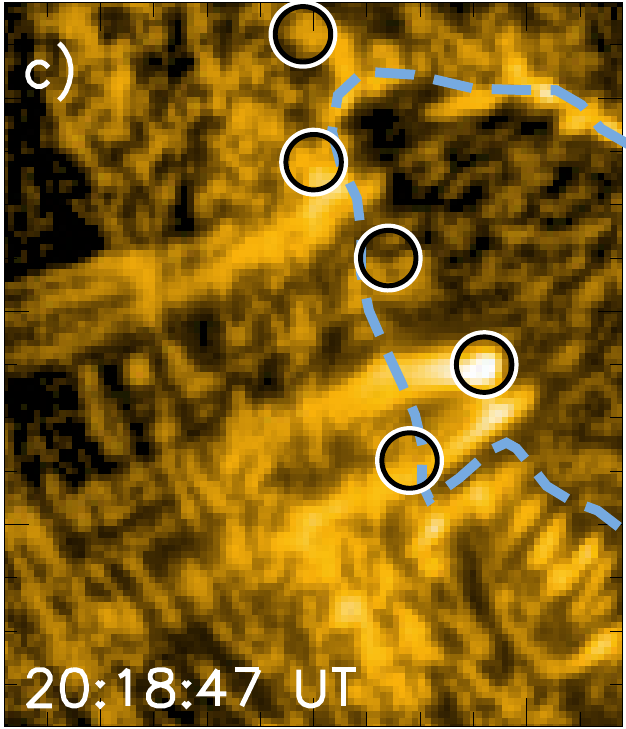}
    \includegraphics[width=3.173cm, clip,  viewport=  98 90 278 300]{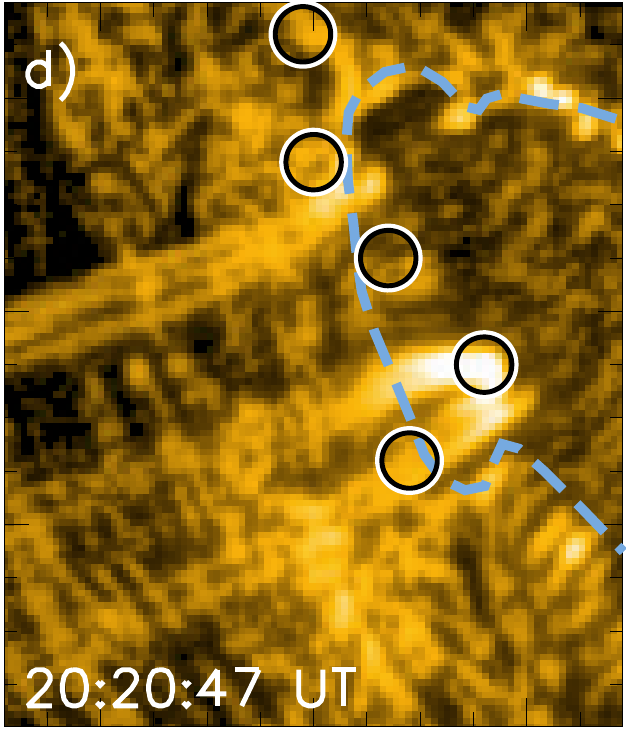}
    \includegraphics[width=3.173cm, clip,  viewport=  98 90 278 300]{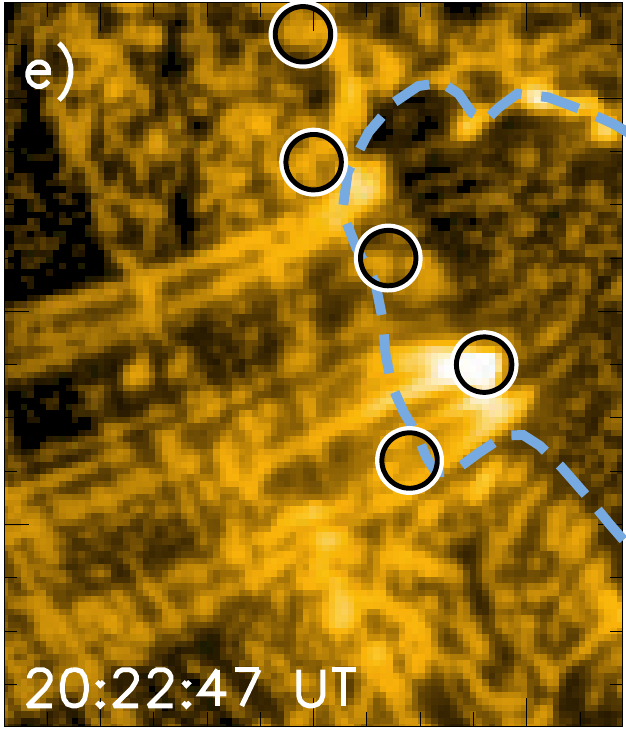}
        \\
    \includegraphics[width=4.9cm, clip,  viewport=  00 30 278 300]{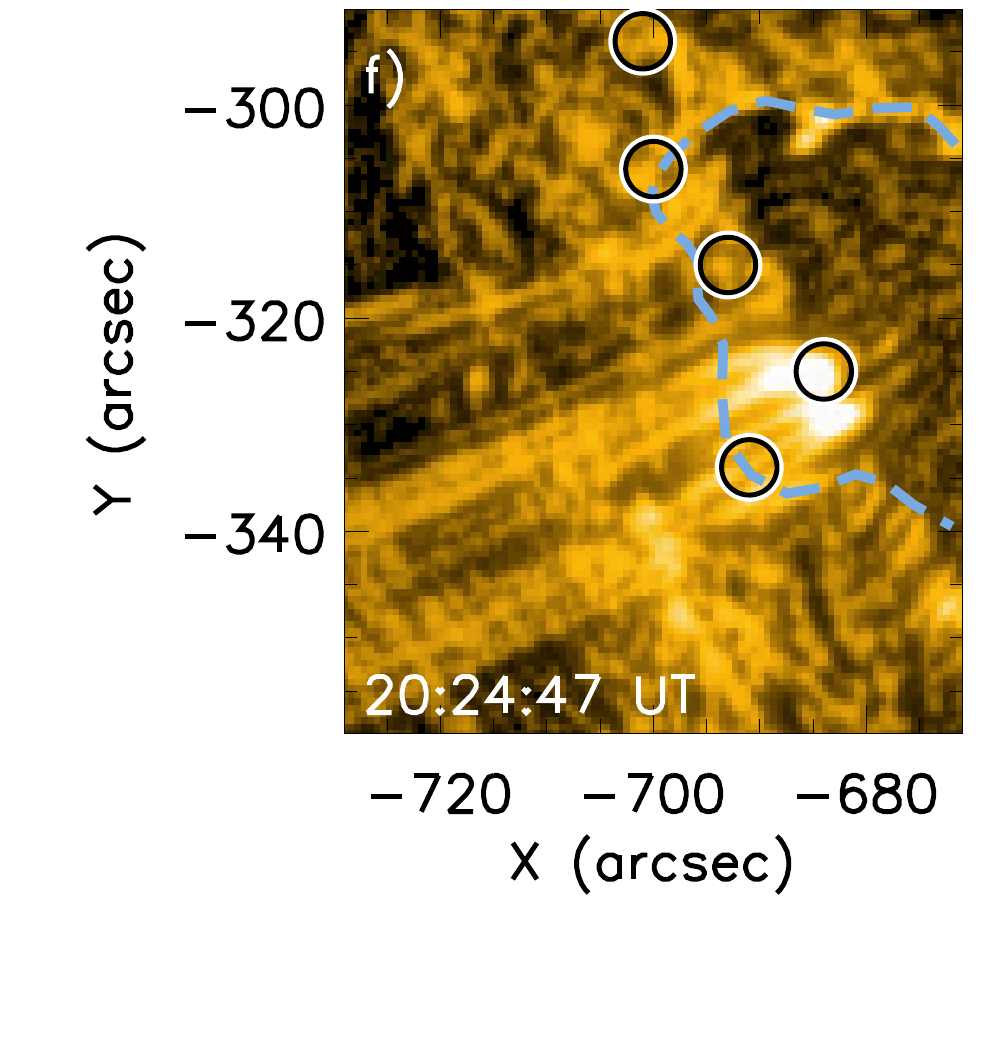}
    \includegraphics[width=3.173cm, clip,  viewport=  98 30 278 300]{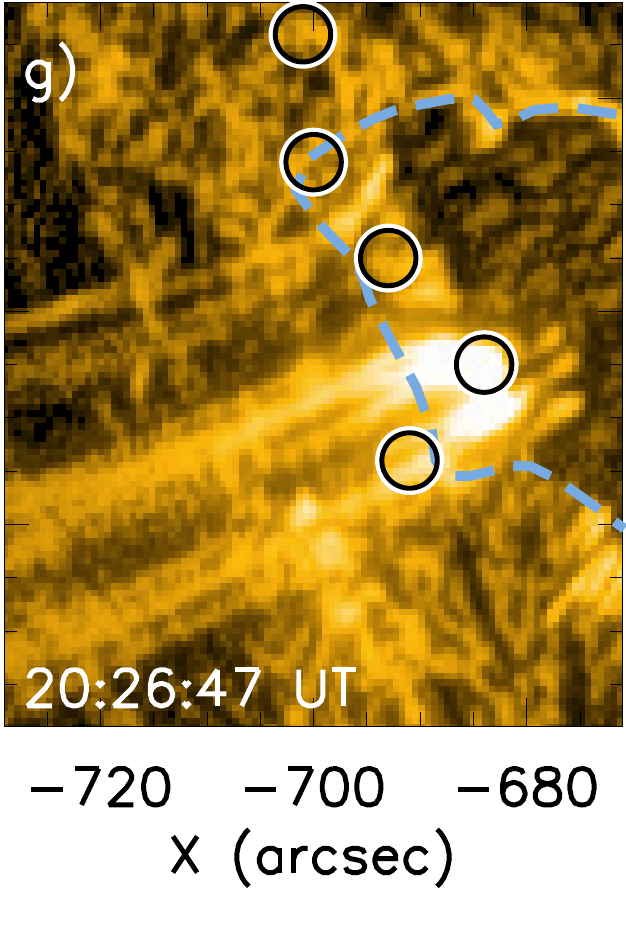}
    \includegraphics[width=3.173cm, clip,  viewport=  98 30 278 300]{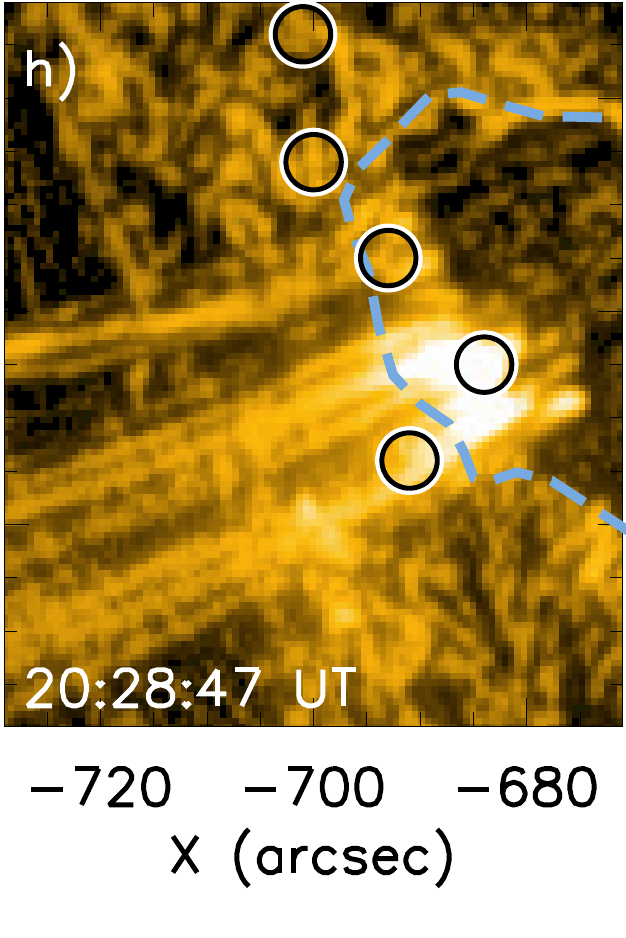}
    \includegraphics[width=3.173cm, clip,  viewport=  98 30 278 300]{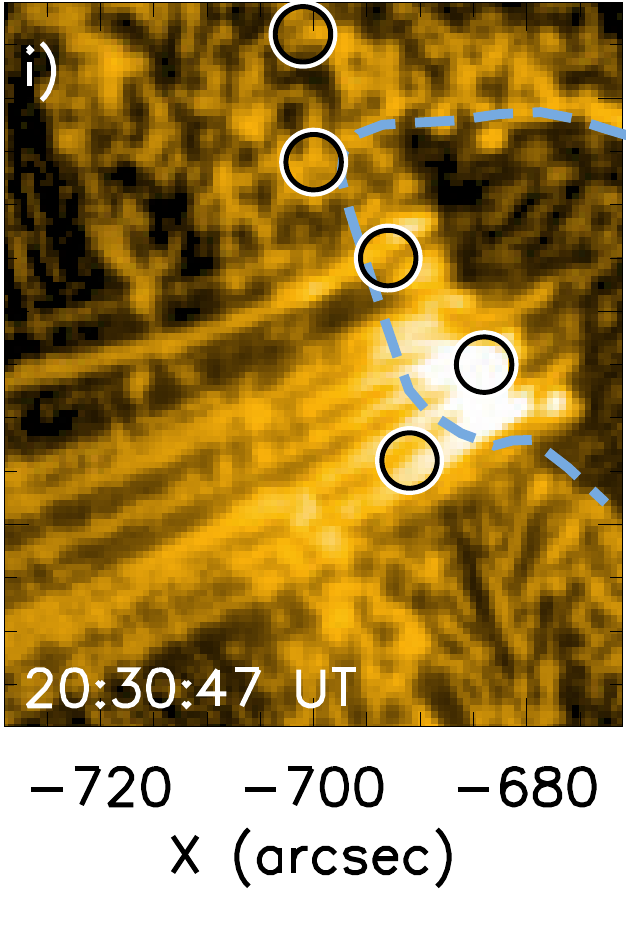}
    \includegraphics[width=3.173cm, clip,  viewport=  98 30 278 300]{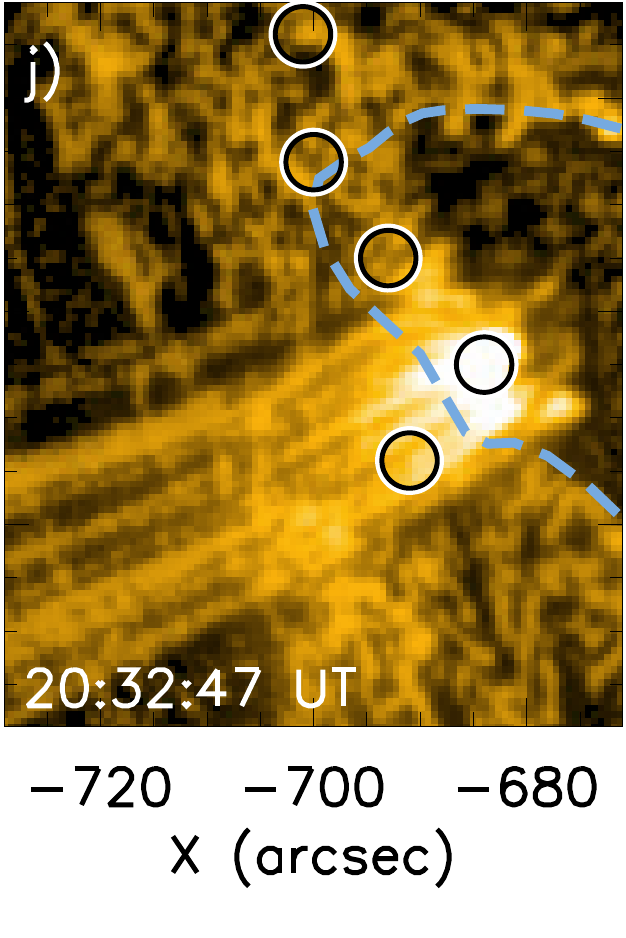}
    \\
    \includegraphics[width=8.cm, clip, viewport=  0 0 467 330]{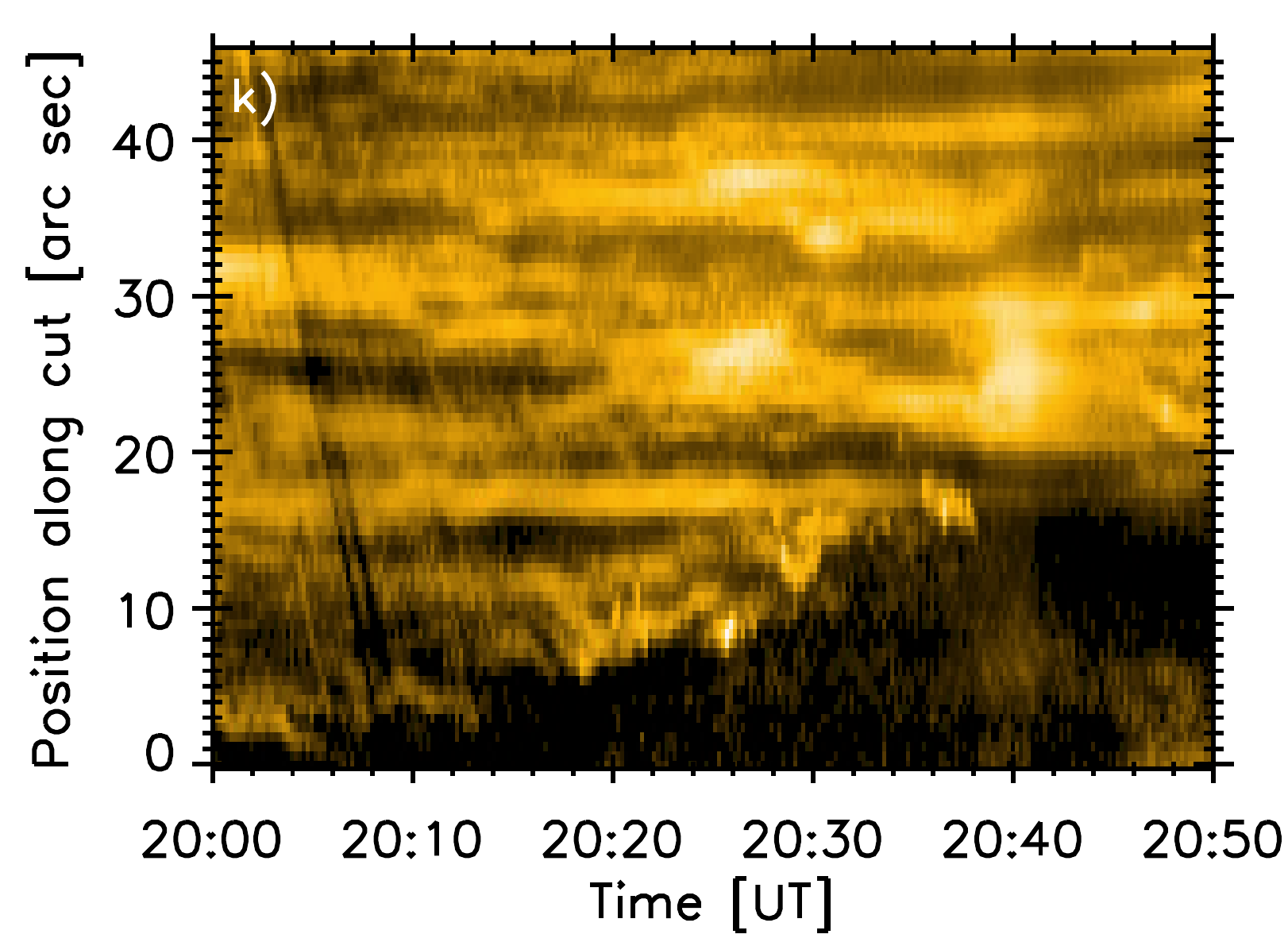}
    \includegraphics[width=8.cm, clip, viewport=  0 0 467 330]{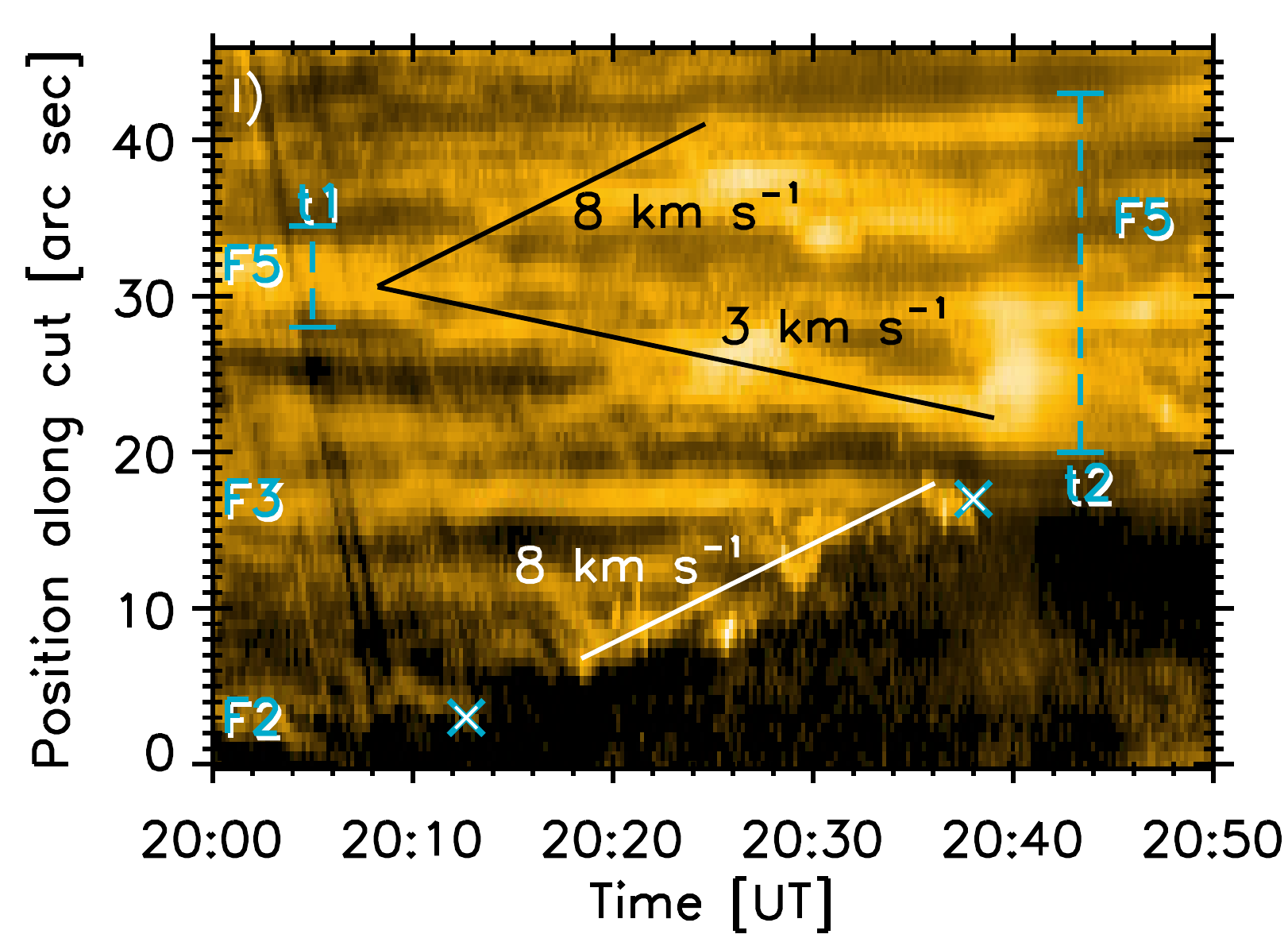}    
	\caption{Panels a--j: detailed view of the filament footpoints in the 171\,\AA{} filter channel data processed with MGN. Circles mark footpoints FR1--FR5, blue dashed lines mark positions of the ribbon found using ratios of the 1600\,\AA{} and 1700\,\AA{} filtergrams. White arrow in panel a) denotes the cut, along which the time-distance diagrams shown in panels k) and l) were constructed. Lines in panel l) denote moving structures. Captions F2 and F3 mark emission of filament strands. Blue 'X' symbols mark the moments when their emission disappeared. Captions F5 and vertical dashed lines mark the filament leg broadening between t1 and t2. Note that the straight inclined dimmings on the left-hand side are caused by the motion of the filament. \\ (An animated version of this figure is available online.) \label{fig_sweep}}
\end{figure*}

\subsection{Sweeping of filament footpoints} \label{sec_drifting}

In this section we focus on the evolution of the filament leg located inside the evolving NRH as well as overlying corona. Figure \ref{fig_nrh_overview} shows NRH and its surroundings as observed in the 171\,\AA{} and 94\,\AA{} filter channels of AIA. 

We first analyze the 171\,\AA{} observations processed using MGN (panels a--l, animated version available in the online journal). There, blue dashed lines mark the derived positions of NRH. At the beginning of the eruption (Figure \ref{fig_nrh_overview}a--b), footpoints of the filament were distributed along a narrow region approximately $\approx$40$\arcsec$ {long}. 

Bright strands composing the filament (hereinafter strands) were found to have footpoints located inside the hook {(black and white cirles named F1--F4). This finding is in agreement with results of 3D simulations of flux rope eruption of \citet{zucc15} and analysis of reconnection geometries occuring at ribbon hooks of \citet{aulanier19}.} According to the HMI observations of the photospheric magnetic field strength, {these footpoints were located in regions of varying \mbox{$B_{\text{LOS}}$} at the boundary of strong-field regions dominated by the negative polarity (Figure \ref{fig_nrh_overview}m).} 


Appearing of NRH seen in panels a) and b) of Figure \ref{fig_nrh_overview} was followed by several changes in the configuration of the filament strands. As NRH started to propagate and contract, the curved part of the hook ('elbow') crossed the footpoint F1 (panel c). {This means that the intensity enhancements corresponding to the chromospheric or transition region flare ribbons changed position, apparently moving through the location of F1 in the plane of sky. \citep[In the following, we will further refer to the locally perpendicular motion of the ribbon as 'sweeping', similarly as in][]{aulanier19}. After the ribbon swept F1, the strands anchored there disappeared (panels c--d). We note that this description is of morphological evolution only, while the physical interpretation is left to Section \ref{interpr3d}.}

At the same time, the footpoint F4 and strands anchored therein started to fade away and drift towards NE. There, they got obscured by new bright strands which appeared in position F5 located inside NRH in a strong-field region \mbox{($B_{\text{LOS}} \approx$--120 G)} close to F3 and F4. Later on, the elbow of the hook propagated towards SW and swept footpoints of strands located between F1 and F2 (panels d--f) and then the footpoints F2 ($\approx$ 20:10 UT, panel f) and F3 ($\approx$ 20:36 UT, panel i). In the meantime, more and more strands anchored in F5 appeared. This trend of footpoint sweeping remained throughout the whole propagation of the hook elbow towards SW, until $\approx$21:00 UT, when all the footpoints were concentrated in F5. Finally, F5 was swept away after $\approx$21:30 UT and all the strands rooted therein disappeared (see panels k--l and the accompanying animation). 

Strands rooted in footpoints F2, F3, F5, and in between them are also detailed in Figure \ref{fig_sweep} (also see the accompanying animation). Panels a--j show a detailed view of NRH at a higher cadence of two minutes. Panels k) and l) show time-distance diagrams containing the emission of filament strands anchored in studied footpoints, constructed along the arrow shown in panel a). The arrow was directed perpendicularly to the filament strands and drawn in a location, where contribution from the coronal moss emitting in the background was low. In panel l), the emission of loops anchored in the footpoints F2, F3, and F5 is denoted by blue captions. Times, at which the emission of strands anchored in F2 and F3 disappeared are marked with blue 'X' symbols. These were found to correspond to the moments at which NRH swept the footpoints of the respective strands, i.e. $\approx$20:14 UT for F2 and $\approx$20:38 UT for F3. In the same panel, a motion of strands lasting between $\approx$20:18 UT -- 20:36 UT can also be seen between the two 'X' symbols. It corresponds to sequential disappearing of filament strands rooted between F2 and F3 due to the sweeping of their footpoints and its velocity was found to be about 8 km\,s$^{-1}$. We note that fitting this motion with a single linear regression was difficult due to bright material downfalling along the strands. This phenomenon occus along majority of the observed strands and is most evident after the onset of the eruption at $\approx$19:30 UT and later between $\approx$20:20 UT and 20:40 UT, i.e. when the footpoints between F2 and F3 were swept (see the animation accompanying Figure \ref{fig_nrh_overview}).  

Broadening of the patch of filament strands anchored in F5 can also be traced in these time-distance diagrams. At $\approx$20:00 UT, the patch of strands anchored in the newly-formed footpoint F5 was about $\approx$6$\arcsec$ wide (t1 in panel l). At about 20:12 UT, after NRH swept F2 and started to sweep the footpoints between F2 and F3, the patch of strands started to broaden in both directions along the cut at uneven velocities of \mbox{8 km\,s$^{-1}$} and \mbox{3 km\,s$^{-1}$}. This apparent drifting lasted for about 30 minutes and in the end, the patch was $\approx$23$\arcsec$ wide (t2 in panel l). 

We note that the broadening of F5 might be caused by the dynamics of the filament plasma itself. Motions of plasma within a stationary filament model were found by \citet{luna2012}. There, authors performed spectral synthesis of condensations moving along filament field lines. The synthetic H$\alpha$, 171\,\AA{}, and 211\,\AA{} emission showed that vertical filament strands can become illuminated by falling filament plasma \citep[see animation b) accompanying Figure 11 in][]{luna2012}, which is similar to our observations. Nevertheless, even the footpoint F5 was eventually swept by the ribbon.

Finally, we note that the period during which the strands rooted in the footpoints F1--F3 seemingly drifted to F5 corresponds well to the impulsive and peak phases of the flare (shaded area in Figure \ref{fig_overview}i).

\subsection{Fading of coronal loops}

\begin{figure}[h]
	\centering
    \includegraphics[width=8.5cm, clip, viewport=  0 0 467 330]{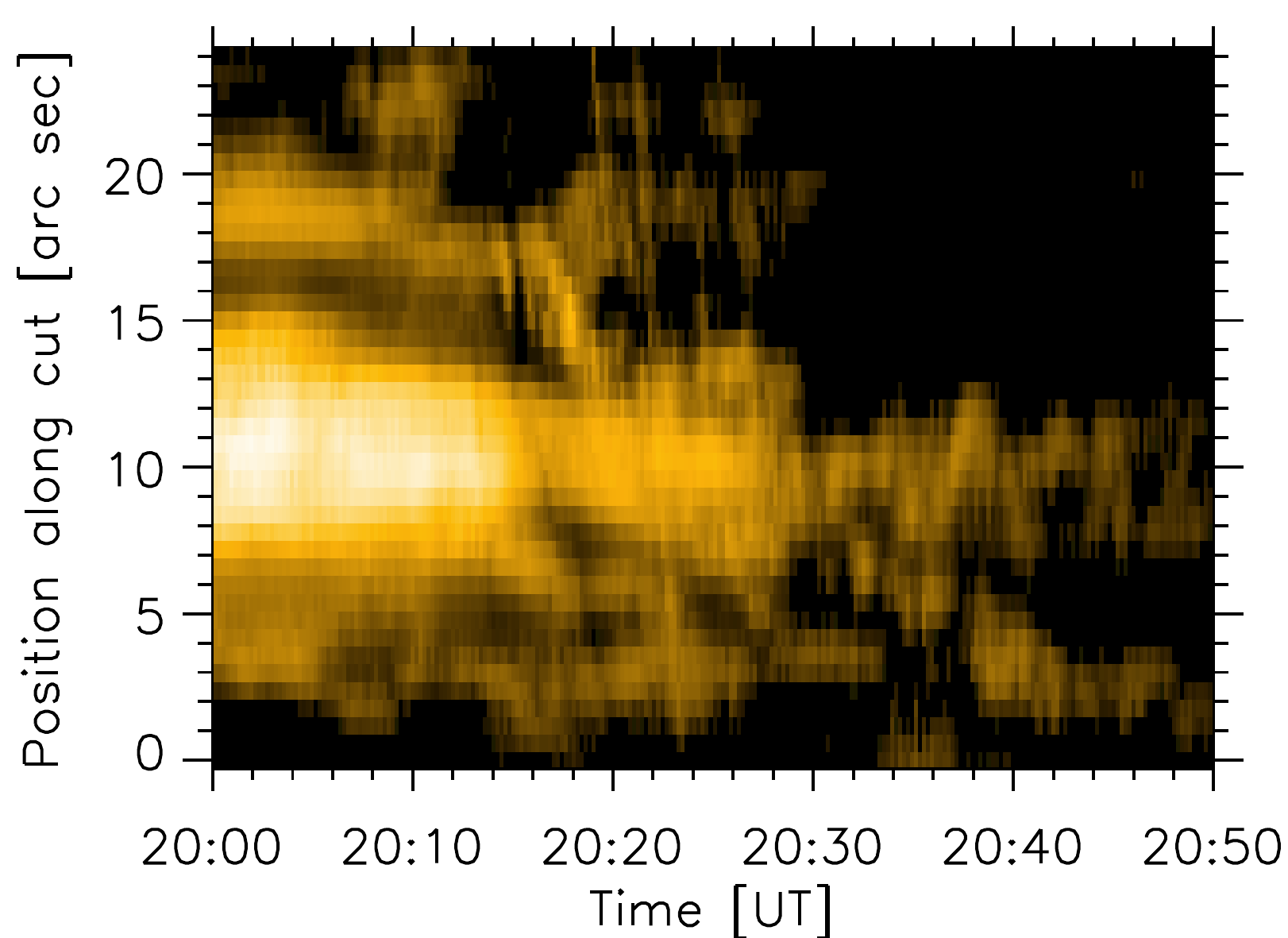}    
	\caption{Time-distance diagram which displays fading of the coronal loops located at the tip of NRH, constructed along the cut shown in Figure \ref{fig_nrh_overview}b. Here, emission of the coronal background was saturated. \label{fig_xt_cl}}
\end{figure}

Figure \ref{fig_nrh_overview} shows a bundle of coronal loops anchored in a neighborhood of NRH ('CL', white arrow in Figure \ref{fig_nrh_overview}a). Figure \ref{fig_nrh_overview}d-e shows that after the tip of NRH elongated in a $V$-shaped trajectory (Section \ref{sec_nrh}), it started to converge towards CL. Later on (panel f), CL became encircled by the tip of NRH and consequently started to fade away as it was swept by NRH (panels f--h). After $\approx$20:36 UT (panel i), CL was not visible anymore. Finally, tip of NRH straightened over the region where CL was located. This can, however, only be seen in the shape of NRH in 171\,\AA{} (panels k--l), as it was not sufficiently bright in the ratios of the 1600\,\AA{} and 1700\,\AA{} filter channels.

The fading of CL was further investigated using time-distance diagram constructed along the cut shown in Figure \ref{fig_nrh_overview}b (Figure \ref{fig_xt_cl}). Here, the background emission was excluded by setting the minimal intensities plotted higher than the intensity of the background (100 DN s$^{-1}$ px$^{-1}$) found by averaging in regions close to CL. Figure \ref{fig_xt_cl} thus only contains the emission of CL and bright filament blobs and strands passing through the FOV during the eruption. According to this time-distance diagram, CL was fading since $\approx$20:04 UT, until it disappeared at $\approx$20:30 UT. The emission present later at position $\approx$10$\arcsec$ of the cut originates in the tip of NRH which appeared under the cut.

Note that the fading of CL is not sequential as is observed in Figure \ref{fig_sweep} for the disappearance of bright filament strands, mainly because of its brief apparent broadening after $\approx$20:16 UT. This effect is caused by the passage of strands of the erupting filament throughout the FOV. Analysis of fading of CL is also complicated by oscillations of CL throughout the course of the eruption (see the animation accompanying Figure \ref{fig_nrh_overview}). Despite these complications, Figure \ref{fig_xt_cl} indicates that the period during which CL faded and disappeared roughly corresponds to the period of sweeping of the filament strands anchored in F2, F3, and between them.

\section{Observations of flare loops} \label{sec_flareloops}

\subsection{Formation of the flare arcade}

To investigate the appearance of flare loops in a vicinity of NRH, we reviewed the 94\,\AA{} filter channel observations (Figure \ref{fig_nrh_overview}n--x, see also the accompanying animation). These filtergrams were averaged in three consecutive exposures to minimize the noise in the 94\,\AA{}. NRH was indicated using orange dashed lines. 

The first flare loops were observed around 19:48 UT and were sheared (Figure \ref{fig_nrh_overview}o, Figure \ref{fig_overview}g). Afterwards, arcade of less-sheared flare loops with footpoints in the elbow of NRH appeared (panels p--q). It was generally difficult to distinguish individual flare loops within this arcade. Some of the conspicuous ones are denoted by 'fl1' and 'fl2' (see panels r) and t), respectively) , however, their footpoints cannot be easily distinguished due to overlapping structures. Later on, as the arcade cooled, flare loops were seen in 171\,\AA{} and 211\,\AA{} filter channels showing cooler emission (Figure \ref{fig_nrh_overview}j--l and Figure \ref{fig_longloop}a, respectively).

As is apparent in panels p--r of Figure \ref{fig_nrh_overview}, most of the arcade formed suddenly without showing any sequence in appearance of flare loops. Therefore, it was not possible to study the evolution of particular flare loops at the same time as the hook swept footpoints of the filament strands. On the other hand, panels q--x show that the footpoints of the arcade correspond to the propagating NRH, which means that flare loops emerge in locations which were originally located inside NRH and then swept by it. Therefore, we emphasize that this arcade is not a result of the standard aa--rf reconnection between two coronal loops, as described in the standard CSHKP model, but occurs in a purely three-dimensional magnetic reconnection geometry.

We note that we also observed emission projected to the interior of NRH, such as flare loops anchored in NRH slipping towards the tip of the hook \citep[orange arrow 'i1' in panel n, see also ][]{lorincik19}, projected flare loops with conjugate footpoints located to the east (arrow 'i2' in panel q), and multithermal emission of filament strands during the broadening of the patch of strands rooted in F5 (arrow 'i3' in panel t).

\subsection{Identification of individual flare loops} \label{sec_cooling}

Because of the nature of the flare loop emission observed in the 94\,\AA{} filter channel, we could not directly identify flare loops formed as a product of reconnection of the filament's bright strands. Therefore, we attempted to find such a flare loop using filter channels typically sensitive to cooler emission such as the 171\,\AA{}, 193\,\AA{}, and 211\,\AA{}. In these filter channels, due to high count rates in active regions, MGN can be utilized in order to enhance individual flare loops within the flare loop arcade.

\begin{figure}[h]
	\centering
	\includegraphics[width=8.5cm, clip,  viewport= 35 25 283 230]{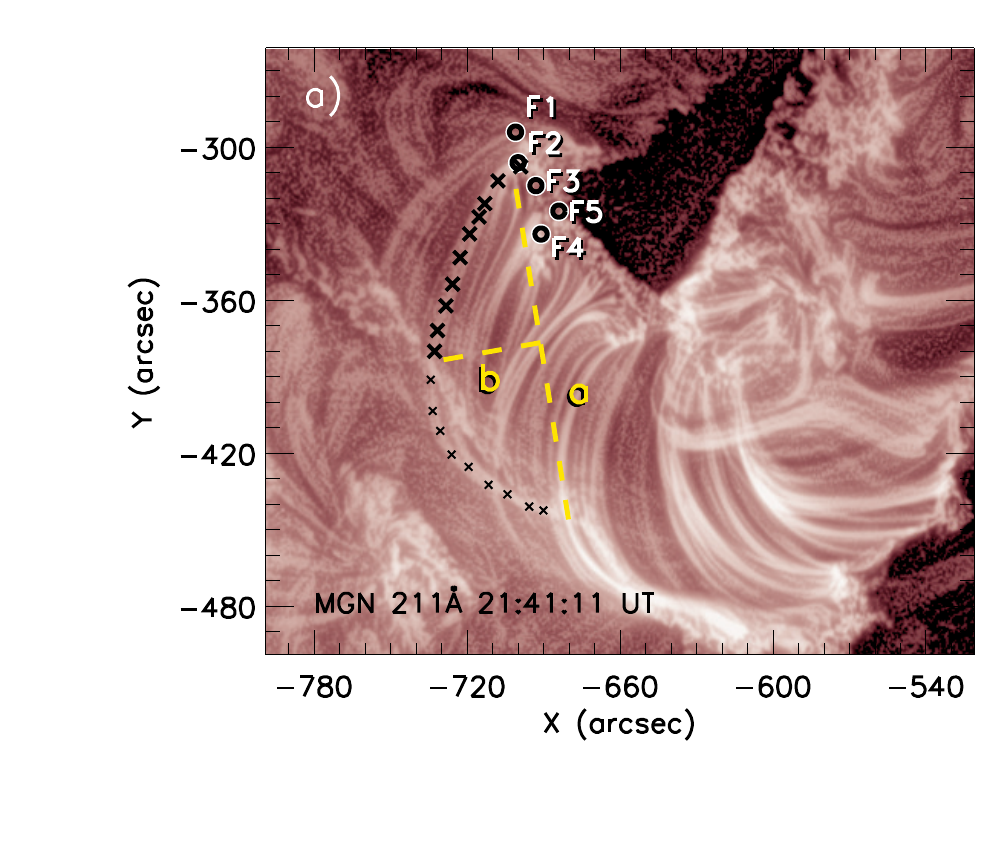}
    \includegraphics[width=8.5cm, clip,  viewport= 0 15 396 283]{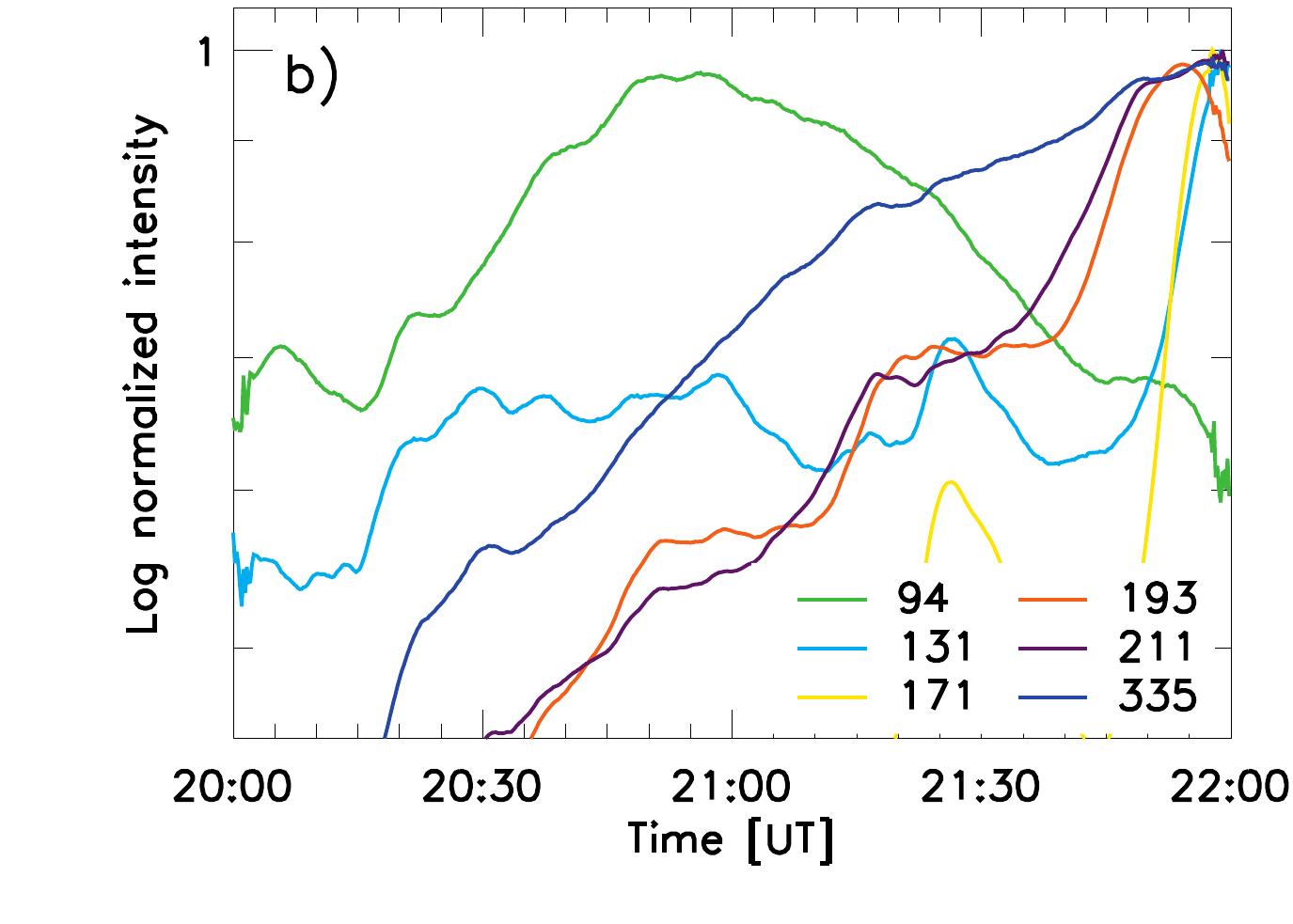}
	\caption{a) AIA 211\,\AA{} observations of the flare arcade. Symbols 'X' mark the selected flare loop. Yellow dashed lines 'a' and 'b' represent major and minor axes of the ellipse used for estimating the length of the selected flare loop. b) Lightcurves of the selected flare loop averaged in positions marked with thick 'X' symbols in panel a) as a function of time. Individual intensity curves were smoothed, normalized, and color-coded in order to distinguish between the filter channels of AIA. \label{fig_longloop}}
\end{figure}

Figure \ref{fig_longloop}a shows the 211\,\AA{} observations of the flare arcade during the gradual phase of the flare. The temperature response of this filter channel peaks at $\approx$2 MK and the emission typically observed in active regions is dominated by \ion{Fe}{14} \citep{delzanna13}. The arcade observed within this channel formed at around 21:00 UT by cooling of the arcade observed in the 94\,\AA{}. Within this arcade, numerous individual flare loops connecting PR with NR and NRH can be seen. In 211\,\AA{}, these loops are bright and their structure is further enhances by the MGN technique. Some of the loops that stand out are anchored in F1, F2, and in F4, i.e. in the previous footpoints of bright strands of the erupted filament (see Figure \ref{fig_nrh_overview}a--b). 

To study the evolution of the flare emission, we selected a distinct loop anchored in F2 ('X'-symbols in Figure \ref{fig_longloop}a) and averaged intensities along it for each EUV filter channel of AIA. Note that for averaging, we only used the positions marked with thick symbols, since the thin ones also include emission of the ribbon PR. Smoothed lightcurves normalized to the maximum of the loop emission are shown in Figure \ref{fig_longloop}b. The emission at the selected locations is first visible in the 94\,\AA{} filter channel. As more and more flare loops appeared, the lightcurve gradually rised to its peak between $\approx$20:45--21:00 UT. After that the arcade started to cool, as the emission increased in the 335\,\AA{} which peak temperature response is at 3 MK and then in the 171\,\AA{}, 193\,\AA{}, and 211\,\AA{} filter channels (1--2 MK). Lightcurves of these filter channels peaked between $\approx$21:50--22:00 UT.

Note that the 131\,\AA{} emission in these locations is weak {, reaching only about 10 \,\dn,} therefore it is likely that the flare loops here do not reach the 11 MK temperatures required for emission {of} \ion{Fe}{21} \citep{dwyer2010,petkaki12}. 


\subsection{Loop cooling timescales}

In order to verify whether the flare loop selected using the 211\,\AA{} data can be associated with the hot emission observed at earlier times, we estimated its cooling time. To do that, we use the combined radiative-conductive cooling time \citep[Equation 14E in ][]{cargill14}:

\begin{equation}
	\tau_{\text{cool}}=\frac{2-\alpha}{1-\alpha}3k_{\text{B}}\left(\frac{1}{\kappa_{\text{0}}^{4-2\alpha}\chi^7}\frac{L^{8-4\alpha}}{(n_{\text{0}}T_{\text{0}})^{3+2\alpha}}\right)^{1/(11-2\alpha)}.
\end{equation}

There, $L$ is the loop length, $n_{\text{0}}$,$T_{\text{0}}$ electron density and temperature, $\kappa_{\text{0}}$ is the Spitzer coefficient of thermal conduction, and $k_{\text{B}}$ is the Boltzmann constant. $\alpha$= --1/2 and $\chi$=--31.5 Wm$^{-1}$K$^{-7/2}$ are parameters of fits of the radiative-loss function \citep[see e.g.][]{dudik11}. 

Loop length was calculated using elliptical approximation of the loop arc. Major and minor axes of the adopted ellipse are indicated with yellow dashed lines in Figure \ref{fig_longloop}a. Both axes of this ellipse were spherically deprojected, assuming that the loop arc is normal to the solar surface. Length of the loop was then found to be $\approx$130 Mm. As we do not posess any spectroscopic observations and loop cooling times are much more dependent on the loop lengths than on $n_{\text{0}}$ and $T_{\text{0}}$, for purposes of rough estimate of $\tau_{\text{cool}}$, we assumed typical electron density in flare loops of $n_{\text{0}}=10^{11}$ cm $^{-3}$ \citep[see e.g.][]{polito17}. Using the aforementioned equation we calculated the time needed for cooling of a loop with an initial temperature of $T$=7.10$^6$K (94\,\AA{}) down to $\approx$2.5$\times10^6$K (211\,\AA{}) as around 85 minutes. Results of calculations of the loop cooling times are also dependent on adopted values of parameters of fits of the radiative loss function. For example, using $\alpha$= --1/2 and $\chi$=--31.81 Wm$^{-1}$K$^{-7/2}$ \citep{kuin82} would result in prolonging of the loop cooling times by about 15$\%$. However, even this result would still correspond, within tens of per cent, to the cooling of plasma indicated by the temporal shifts between the maxima of lightcurves in Figure \ref{fig_longloop}b. Therefore, the selected loop could have been among the hot flare arcade observed in the 94\,\AA{} data after the onset of the eruption. We note that such long cooling timescales are typical for long active region loops \citep[see e.g.][]{li15,froment15,lionello16}. 
 
\section{Discussion and conclusions} \label{sec_conclusions}

\subsection{Summary of observations}

In this work we analyzed evolution of a hook of one of the ribbons observed during the eruption of a quiescent filament observed on 2012 August 31. Before the eruption, bright strands composing the filament were anchored in footpoints F1--F4 located inside the ribbon hook. After the onset of the eruption, propagating hook started to sweep these footpoints, which resulted in disappearance of strands anchored therein and was followed by formation of flare arcade rooted in regions swept by the ribbon hook. This lasted for about 150 minutes, until all the filament strands were swept. Since the apparent boundary between the newly-formed flare loops and the filament strands well corresponded to the propagating hook, this was likely caused by magnetic reconnection.

In the following, we provide a brief summary of the evolution of each footpoint and filament strands rooted therein. 

\begin{itemize}

\item[{F1:}] At the onset of the eruption, the ribbon hook appeared close to the northern footpoint F1. The hook then started to elongate as well as propagate in perpendicular direction and swept F1. As a consequence, strand anchored in this footpoint disappeared. A few minutes later, flare loop with a footpoint located in a neighborhood appeared in the 94\,\AA{} channel.

\item[{F2 -- F3:}] Subsequently, the propagating hook swept the footpoints F2 and F3, which then became footpoints of flare loops. At the same time, we observed fading of a bundle of long and straight coronal loops CL located at the tip of the hook. 

\item[{F4:}] Strands anchored in the footpoint F4 started to drift and fade away at the onset of the eruption. Their motion was traced only for a brief period of time, as they eventually became obscured by strands anchored in the position F5. Nevertheless, F4 also became footpoint of flare loops, as it got eventually swept by the hook.
 
\item[{F5:}] The footpoint F5 and strands anchored therein appeared roughly at the same time as the hook swept F1. In the animation accompanying Figure \ref{fig_nrh_overview} it can be seen that the first strands which appeared in F5 could have been formed by downfalling material from larger altitudes. Therefore, field lines composing the erupting flux rope might have been present at this postion even before the eruption of the filament, but only became visible after being illuminated with emitting material. Later on, during the sweeping of the footpoints F2--F3, the patch of strands rooted in F5 widened. F5 was, however, eventually swept by the hook too and strands rooted therein turned into flare loops. 

\end{itemize}

The arcade of hot flare loops was observed in the 94\,\AA{} filter channel in regions swept by the hook. Due to fuzziness of the arcade in this filter channel, we could not distinguish individual flare loops within the arcade. However, well-defined flare loops were observed later on in the coronal channels of AIA. Some of them were found to originate in the swept footpoints F1--F5. By estimating the cooling time in one particular case, we verified that these loops correspond to the hot 94\,\AA{} arcade observed after the onset of the eruption.

\subsection{Interpretation in Terms of 3D Reconnection} \label{interpr3d}

In all five cases listed above, filament strands were observed to turn into flare loops once swept by the propagating ribbon hook. {Propagation of ribbons is, according to the standard CSHKP model of solar flares, consequence of ongoing magnetic reconnection. However}, reconnection of filament field lines into flare loops, {occuring at evolving ribbon hooks,} cannot be described by this model, since it does not include ribbon hooks in which footpoints of erupting flux ropes are located. Therefore, three-dimensional approach is required to address {this observed} conversion of filament strands into flare loops.


{Our} observations are, however, consistent with the recent predictions of the 3D extensions of the standard solar flare model of \citet{aulanier19}. These authors identified new three-dimensional reconnection geometries that involve the flux rope field line, located within the hooks of the QSL footprints, reconnecting to become a flare loop. In one of these reconnection geometries, the ar--rf reconnection, flux rope field lines composing a modelled flux rope (\textit{r}) reconnect with coronal arcades (\textit{a}) into new field lines composing the flux rope (\textit{r}) and flare loops (\textit{f}). 

These predictions are in our case consistent with observations of the disappearance of filament threads and formation of flare loops (i.e. 'r' $\rightarrow$ 'f'). Moreover sweeping of footpoints F2--F3 was found to be contemporal with gradual fading of coronal loops CL which could be the 'a' component of the ar--rf reconnection. If so, the previous location of the CL footpoint should become a part of the post-reconnection flux rope 'r'. This is the case here. Even though no particular flux rope field line is seen to originate from this location after it is swept by the hook, it is now a part of coronal dimming area, which are interpreted as consequences of evacuation of flux rope plasma by the eruption \citep[e.g.,][]{dissauer18}.

However, note that unlike the model, where individual field lines can be traced in their entirety, the observations only show lower parts of filament strands and lower parts of coronal loops rooted nearby. This is due to the erupting filament being large, with length comparable to the solar radius (Figure \ref{fig_overview}a--b). We were not able to identify the conjugate footpoints of CL near the positive-polarity ribbon hook and thus verify if they composed an arcade overlying the filament (\textit{a}), as predicted by the model. Nevertheless, the observed dynamics of lower portions of the filament strands being swept by the ribbon together with the coronal loops, conforms well with the predictions of the model.

Despite the MHD model used in \citet{aulanier19} is in the zero-$\beta$ approximation and thus it does not predict the emission of the constituents of individual reconnection geometries, we have for the first time shown that the filament strands reconnect to become flare loops. This indicates a 3D reconnection process irrespective of the fact that some constituents of the ar--rf reconnection geometry were not identifiable in the present filament eruption. In the future, more events, especially eruptions of filaments with bright and distinguished footpoints, need to be studied for identification of all four constituents partaking in the ar--rf and other reconnection geometries predicted by \citet{aulanier19}.

\subsection{Conclusions}

In summary, we observed sweeping of many filament footpoints by the propagating hook, which was always followed by appearance of flare loops rooted in the original filament footpoints. This reconnection of field lines composing the filament into flare loops is not described by the standard CSHKP model of solar flares, since this model does not include the hooked ribbons. {It however conforms to the predictions of the 3D extensions to the standard model as described in \citet{aulanier19}, where flux rope field lines at the hook can reconnect to become flare loops. In our observations, this phenomenon was observed for all of the visible filament footpoints located within the hook.}

In addition, the sweeping of some of the filament strands and footpoints was found to be cotemporal with disappearance of a nearby coronal arcade. These observations are consistent with the three-dimensional ar--rf reconnection geometry predicted by \citet{aulanier19}, in which field lines of the modelled flux rope reconnect with a coronal arcade at the ribbon hook into new field lines composing the flux rope and a flare loop. Since only three constituents of the ar--rf reconnection geometry were clearly identifiable in the present filament eruption, more events, especially eruptions of filaments with bright and distinguished footpoints, should be studied for identification of all four constituents partaking in this reconnection geometry.

\vspace{1cm}
{The authors thank the referee for comments that helped to improve the manuscript. The} authors {also} thank Brigitte Schmieder for useful discussions. J.L. and J.D. acknowledge the project 17-16447S of the Grant Agency of Czech Republic as well as insitutional support RVO: 67985815 from the Czech Academy of Sciences. G.A. thanks the CNES and the Programme National Soleil Terre of the CNRS/INSU for financial support, as well as the Astronomical Institute of the Czech Academy of Sciences in Ond\v{r}ejov for financial support and warm welcome during his visit. AIA and HMI data are provided courtesy of NASA/SDO and the AIA and HMI science teams.
\\
\textit{Facilities:} SDO, Hinode

\bibliographystyle{aasjournal}
\bibliography{arrf_submit}

\end{document}